\documentclass[a4paper,fleqn]{cas-sc}

\usepackage[authoryear,longnamesfirst]{natbib}
\usepackage{subfigure}
\usepackage{algorithm,algpseudocode,multicol}
\usepackage{soul} 
\usepackage{xcolor}
\usepackage{diagbox}
\newcommand\emptyDiag[2][]{
	\diagbox[innerwidth=\widthof{#2},height=\line, #1]{}{}%%
}

\def\tsc#1{\csdef{#1}{\textsc{\lowercase{#1}}\xspace}}
\tsc{WGM}
\tsc{QE}
\tsc{EP}
\tsc{PMS}
\tsc{BEC}
\tsc{DE}
\begin{document}
	%\begin{sloppypar}
%\let\WriteBookmarks\relax
%\def\floatpagepagefraction{1}
%\def\textpagefraction{.001}

%
%\theoremstyle{plain}
\newtheorem{thm}{Theorem}
\newtheorem{lem}{Lemma}
\newtheorem{cor}{Corollary}
\newtheorem{rem}{Remark}
\newtheorem{assum}{Assumption}
\newtheorem{prob}{Problem}
\newtheorem{defn}{Definition}
\newproof{pf}{Proof}
\newproof{pot}{Proof of Theorem \ref{thm2}}

\shorttitle{Robust Set Partitioning Strategy for Malicious Information Detection in Large-Scale Internet of Things}

\shortauthors{Suo et~al.}

\title [mode = title]{\textcolor{blue}{Robust Set Partitioning Strategy for Malicious Information Detection in Large-Scale Internet of Things}}
%{How to Divide: A Set Partitioning Strategy Balancing the Trade-off Between Intra-Subset Correlation and Inter-Subset Gain Mutual Influence in Distributed Attack Detection Scheduling Task}                      
%\tnotemark[1,2]
%
%% Title footnote 1.
%% eg: \tnotetext[1]{Title footnote text}
%% \tnotetext[<tnote number>]{<tnote text>} 
%\tnotetext[1]{This document is the results of the research
%   project funded by the National Science Foundation.}
%
%\tnotetext[2]{The second title footnote which is a longer text matter
%   to fill through the whole text width and overflow into
%   another line in the footnotes area of the first page.}


%\author[1]{***}
%\affiliation[1]{***}

\author[1]{Yuhan Suo}[style=chinese]
%[type=editor,
%                        auid=000,bioid=1,
%                        prefix=Sir,
%                        role=Researcher,
%                        orcid=0000-0001-7511-2910]



\ead{yuhan.suo@bit.edu.cn}
\credit{Conceptualization, Methodology, Software, writing-original draft}
%\credit{Data curation, Writing - Original draft preparation}
%\ead[url]{www.cvr.cc, cvr@sayahna.org}

%\credit{Conceptualization of this study, Methodology, Software}

\affiliation[1]{organization={School of Automation},
    addressline={Beijing Institute of Technology}, 
    city={Beijing},
    % citysep={}, % Uncomment if no comma needed between city and postcode
    postcode={100081}, 
    country={China}}

\author[1]{Runqi Chai}[style=chinese,orcid=0000-0003-4083-8863]
\ead{r.chai@bit.edu.cn}
\credit{Methodology, Writing - Review \& Editing}
\cormark[1]

\author[2,3]{Kaiyuan Chen}[style=chinese]
\ead{kaiyuanchen@mail.tsinghua.edu.cn}
\credit{Writing - Review \& Editing}
\cormark[1]

\affiliation[2]{organization={Vanke School of Public Health, Institute for Healthy China},
	addressline={Tsinghua University}, 
	city={Beijing},
	% citysep={}, % Uncomment if no comma needed between city and postcode
	postcode={100084}, 
	country={China}}
	
\affiliation[3]{organization={The State Key Laboratory of Management and Control
		for Complex Systems, Institute of Automation},
		addressline={Chinese Academy of Sciences}, 
		city={Beijing},
		% citysep={}, % Uncomment if no comma needed between city and postcode
		postcode={100190}, 
		country={China}}
	
\author[1]{Senchun Chai}[style=chinese]
\ead{chaisc97@bit.edu.cn}
%\ead[URL]{www.sayahna.org}
\credit{Review, supervision}

\author[2]{Wannian Liang}[style=chinese]
\ead{liangwn@
	tsinghua.edu.cn}
%\ead[URL]{www.sayahna.org}
\credit{Writing - Review \& Editing}
%\affiliation[2]{organization={Sayahna Foundation},
%    % addressline={}, 
%    city={Jagathy},
%    % citysep={}, % Uncomment if no comma needed between city and postcode
%    postcode={695014}, 
%    state={Trivandrum},
%    country={India}}
%\affiliation[3]{organization={Key Laboratory of Fieldbus Technology
%		and Automation of Beijing},
%	addressline={North China University of Technology}, 
%	city={Beijing},
%	% citysep={}, % Uncomment if no comma needed between city and postcode
%	postcode={100144}, 
	%country={China}}
	
%\author[4]{Jiping Xu}[style=chinese]
%%\fnmark[2]
%\ead{xujp@th.btbu.edu.cn}
%%\ead[URL]{www.sayahna.org}
%\credit{Review}
%% Address/affiliation
%%\affiliation[2]{organization={Sayahna Foundation},
%	%    % addressline={}, 
%	%    city={Jagathy},
%	%    % citysep={}, % Uncomment if no comma needed between city and postcode
%	%    postcode={695014}, 
%	%    state={Trivandrum},
%	%    country={India}}
%\affiliation[4]{organization={School of Computer and Artificial Intelligence},
%	addressline={Beijing
%		Technology and Business University}, 
%	city={Beijing},
%	% citysep={}, % Uncomment if no comma needed between city and postcode
%	postcode={100048}, 
%	% state={},
%	country={China}}

\author%
[1]
{Yuanqing Xia}[style=chinese]
%\cormark[2]
%\fnmark[1,3]
\ead{xia_yuanqing@bit.edu.cn} 
\credit{Review}
%\ead[URL]{www.stmdocs.in}

%\affiliation[3]{organization={STM Document Engineering Pvt Ltd.},
%    addressline={Mepukada}, 
%    city={Malayinkil},
%    % citysep={}, % Uncomment if no comma needed between city and postcode
%    postcode={695571}, 
%    state={Trivandrum},
%    country={India}}

\cortext[cor1]{Corresponding author}
%\cortext[cor2]{Principal corresponding author}

%% Footnote text
%\fntext[fn1]{This is the first author footnote. but is common to third
%  author as well.}
%\fntext[fn2]{Another author footnote, this is a very long footnote and
%  it should be a really long footnote. But this footnote is not yet
%  sufficiently long enough to make two lines of footnote text.}
%
%% For a title note without a number/mark
%\nonumnote{This note has no numbers. In this paper we demonstrate $a_b$
%  the formation Y\_1 of a new type of polariton on the interface
%  between a cuprous oxide slab and a polystyrene micro-sphere placed
%  on the slab.
%  }

\begin{abstract}
	With the rapid development of the Internet of Things (IoT), the risks of data tampering and malicious information injection have intensified, making efficient threat detection in large-scale distributed sensor networks a pressing challenge. To address the decline in malicious information detection efficiency as network scale expands, this paper investigates a robust set partitioning strategy and, on this basis, develops a distributed attack detection framework with theoretical guarantees. Specifically, we introduce a gain mutual influence metric to characterize the inter-subset interference arising during gain updates, thereby revealing the fundamental reason for the performance gap between distributed and centralized algorithms. Building on this insight, the set partitioning strategy based on Grassmann distance is proposed, which significantly reduces the computational cost of gain updates while maintaining detection performance, and ensures that the distributed setting under subset partitioning preserves the same theoretical performance bound as the baseline algorithm. Unlike conventional clustering methods, the proposed set partitioning strategy leverages the intrinsic observational features of sensors for robust partitioning, thereby enhancing resilience to noise and interference. Simulation results demonstrate that the proposed method limits the performance gap between distributed and centralized detection to no more than 1.648$\%$, while the computational cost decreases at an order of $O(1/m)$ with the number of subsets $m$. Therefore, the proposed algorithm effectively reduces computational overhead while preserving detection accuracy, offering a practical low-cost and highly reliable security detection solution for edge nodes in large-scale IoT systems.

\end{abstract}

% \includegraphics{figs/grabs.pdf}

 %Research highlights
%	%面对更加隐蔽的自传播攻击，创新性地用移动汇聚节点动态跟踪攻击者。
%	%对于不可监控攻击，提出一种能够仅依据部分置信信息的移动策略设计算法。
%	%给出了移动汇聚节点能够在有限次移动后跟踪到任意攻击者的一系列确定条件。
%
%\item 	The detection problem of dynamic propagation attacks in the network is transformed into the problem of tracking dynamic attacks using mobile sink nodes.
%
%\item  The mobile strategy design algorithm based on partial confident information is proposed to track the unmonitorable attack.
%
%\item A series of deterministic conditions are given for the mobile sink node to track any attacker after a limited number of movements.

% Keywords

\maketitle
\setlength{\parindent}{2em} 
\section{Introduction}
\subsection{Background and Related works}
\textcolor{blue}{The rapid development of the IoT is driving digital transformation across key sectors such as energy, power, healthcare, and smart manufacturing. Billions of heterogeneous devices continuously collect and transmit massive volumes of data through sensors and network interfaces, making IoT a fundamental enabler of intelligent control and automation \citep{zhang2020empowering}. }

	\textcolor{blue}{However, large-scale connectivity, device heterogeneity, and long life cycles expose unique attack surfaces across the device–edge–network–cloud continuum, posing unprecedented security challenges. Resource constraints and “headless deployments” complicate patch management, while issues such as default credentials, insecure firmware updates, and expired certificates remain widespread \citep{yousefnezhad2020security}. Lightweight protocols (e.g., MQTT, CoAP), if lacking encryption and key rotation, are highly vulnerable to man-in-the-middle and replay attacks \citep{mathews2019protocol}. At the Information Technology(IT)/Operational Technology(OT) convergence boundary, IoT systems must balance real-time performance with reliability, where security failures can escalate directly into risks at the physical layer \citep{pascoe2023public}. Ensuring system reliability and data trustworthiness in such large-scale, heterogeneous environments has therefore become a central challenge for the widespread adoption of IoT applications \citep{alsalem2023towards}.}

\textcolor{blue}{At the framework and methodology level, several widely adopted cybersecurity frameworks provide foundational guidance for IoT security. NIST CSF defines an organizational risk management cycle through five core functions, includes Identify, Protect, Detect, Respond, and Recover \citep{pascoe2023public,aljumaiah2025analyzing}. IEC 62443, tailored for industrial control and OT systems, highlights layered and defense-in-depth strategies \citep{leander2019applicability}. ISO/IEC 27001 and 30141 offer general guidance on information security management and IoT reference architectures \citep{humphreys2016implementing,sugiharto2023architecture}, while ENISA’s threat landscape reports regularly update major IoT threats and mitigation trends across Europe and beyond \citep{ENISA2018ThreatLandscape}. Despite their value, these frameworks largely operate at a strategic and governance level, with limited emphasis on dynamic attack detection and response mechanisms required in practice.}

\textcolor{blue}{Beyond the general governance frameworks proposed by standardization bodies, the academic community has also explored security frameworks tailored to IoT scenarios. \cite{halgamuge2025adaptive} introduced an adaptive edge security framework capable of dynamically generating security policies to address complex and evolving threats. To tackle security issues arising from IoT’s heterogeneous architectures, \cite{masud2025vulnerability} proposed a hybrid moving target defense (MTD)-based security level analysis method for assessing network states.  \cite{pavithran2020towards} identified key elements for building secure blockchain architectures in IoT, contributing to improved system robustness. Addressing external intrusion risks, \cite{qaddos2024novel} designed a deep learning architecture capable of capturing complex features, offering potential for safeguarding IoT against security threats. Overall, these studies have achieved notable progress in authentication, protection, and anomaly detection, and to some extent help bridge the gap between high-level standardized frameworks and their practical implementation.}

\textcolor{blue}{Sensor networks form a core component enabling collaboration among heterogeneous IoT devices. However, as the scale of IoT systems continues to expand, sensor networks themselves are growing at an exponential rate. This trend introduces complex security and governance challenges, as decision-makers must balance efficient system operation with the timely detection and response to emerging threats \citep{xie2018data}. Adversaries may disrupt state estimation through data tampering or information deception, leading to system malfunctions and even severe security incidents that pose significant risks to critical infrastructure and essential services \citep{ge2019distributed,pang2021false,smith2021user}. Although end-to-end security mechanisms, such as schemes based on pre-shared keys or digital certificates, have been deployed in some sensor networks, the compromise of keys or certificates can still result in data tampering and integrity breaches \citep{kwon2013security,mouha2021review}. Moreover, despite substantial progress in IoT security research, studies on large-scale distributed sensor networks remain limited, particularly with respect to addressing complex integrity attacks that are both dynamic and stealthy. Developing efficient and precise defensive mechanisms has therefore become a critical research direction for ensuring the stability and reliability of such systems.}

\textcolor{blue}{Attack detection has long been a core area of research in cybersecurity,  encompassing critical domains such as intrusion detection, anomaly detection, and fault diagnosis \citep{wang2025locational,suo2024opinion,li2023fast,li2024detecting,zhang2021deep,zhao2023sparse,balta2023digital}. Existing methods primarily fall into two categories: rule-based and learning-based approaches. 
Rule-based approaches rely on predefined security policies and pattern matching techniques, such as anomaly threshold settings and signature detection. For example,  \cite{wang2025locational} extracted topological relationships measured by individual nodes and used reconstruction residuals to pinpoint the location of injection attacks.  For the situation where malicious agents are in a dominant position, \cite{suo2024opinion} explored using latent features to iteratively filter out malicious agents. \cite{li2023fast} constructed a dynamic relationship between raw data and decrypted results to enable rapid detection of malicious attacks. In the realm of manufacturing system security,  \cite{li2024detecting} developed detection rules based on processing procedures and key parameters, identifying network attacks through rule matching. However, rule-based approaches struggle to adapt to novel and complex attacks.
In contrast, learning-based approaches leverage data-driven techniques such as statistical analysis, machine learning, and deep learning to detect and recognize attack patterns, offering superior generalization and adaptability \citep{zhang2021deep}. For instance, \cite{zhao2023sparse} designed a data-driven attack detector based on subspace identification, detecting attacks by computing the system’s stable kernel representation.  \cite{balta2023digital} proposed a detection framework based on digital twins, identifying attacks by analyzing controlled transient behaviors. To address traffic anomaly detection, \cite{zhang2022manomaly} designed a model consisting of two adversarial sub-networks to learn the data distribution of normal traffic. And based on this, an anomaly detection method based on high anomaly suppression is proposed.
Nevertheless, existing research indicates that these methods still face significant challenges in model training and data dependency \citep{li2024detecting,zhang2021deep}. 
Thus, improving adaptability and computational efficiency while ensuring detection accuracy remains a critical research direction in attack detection.}

\textcolor{blue}{Beyond the research of attack detection, enhancing the system’s resilience is also crucial for ensuring stability. Secure state estimation aims to accurately recover the state of system even in the presence of attack. This problem has become a key research focus in the field of cyber-physical system security, particularly in control systems and distributed sensor networks \citep{ding2020secure,mustafa2022secure}.  
In distributed sensor networks, redundant information is considered essential for ensuring secure state estimation. Even when some sensors are compromised, redundancy allows the system to maintain normal operation \citep{shoukry2017secure}. However, identifying compromised sensors efficiently while avoiding the combinatorial explosion associated with NP-hard problems remains a major research challenge \citep{lu2023secure,an2022fast,lu2023polynomial}. 
To enhance system resilience, researchers have proposed various secure state estimation strategies. For instance, when core sensors are compromised,  \cite{xin2025secure} introduced an estimation method based on virtual sensors, integrating deep reinforcement learning for online optimization to improve estimation accuracy and reliability. This approach aligns well with the concept of redundant information.   \cite{xia2025resilient} proposed a robust distributed Kalman filtering algorithm, which leverages attack detection and robust data fusion strategies to mitigate the impact of malicious network attacks, thereby enhancing the accuracy and stability of distributed state estimation.  
However, in large-scale distributed sensor networks, computational complexity remains a significant challenge. Balancing security and computational efficiency continues to be a pressing issue that requires further investigation in this field.}

\subsection{Motivation and Contributions}
\textcolor{blue}{In large-scale sensor networks, each sensor is typically connected to a substantial number of neighbors. In such scenarios, if every sensor has to sequentially identify potential malicious information from its entire neighborhood, the computational overhead becomes prohibitively high and the detection efficiency drops significantly \citep{Suo2024Security}. Therefore, achieving both system security and computational efficiency within local neighborhoods remains a critical and unsolved problem.}

\textcolor{blue}{Recent studies have shown that partitioning large-scale datasets and performing parallel screening across subsets can help mitigate the curse of dimensionality and significantly improve efficiency \citep{wang2023fuzzy}. Existing clustering and partitioning techniques, however, typically rely on the process of first collecting data and then grouping nodes based on correlation or similarity features \citep{xia2021granular, mirzasoleiman2016distributed}. In practice, the collected data are often affected by communication noise and external interference, resulting in unstable classification and suboptimal partitioning outcomes \citep{han2022data}. In particular, under malicious attacks, existing classification algorithms are primarily designed to detect the attacks themselves rather than to pre-classify datasets that already contain adversarial information \citep{ding2022imbalanced,thakkar2023attack}. Motivated by these limitations, this paper seeks to explore more inherent and fundamental data features to enable stable and effective grouping in the offline stage.}

\textcolor{blue}{	At the same time, although distributed methods are inherently more suitable for large-scale scenarios in terms of efficiency, they still suffer from significant performance drawbacks compared to centralized approaches \citep{mirzasoleiman2016distributed}. One key reason is the lack of a systematic understanding of the relationship between partition strategies and node features. To address this issue, this paper further focuses on optimizing partition strategies, with the goal of narrowing the performance gap between centralized and distributed methods after partitioning, thereby achieving a better balance between efficiency and performance in malicious information selection tasks.  The main contributions of this paper are summarized as follows:}

\begin{enumerate}
	\item  \textcolor{blue}{This paper proposes a “mutual influence” metric to quantify the marginal interference among different sensor subsets. This metric provides a quantitative basis for set partitioning, enabling the partition to maximize the intra-subset correlation while minimizing inter-subset interference, thereby revealing the fundamental reason of the performance gap between distributed and centralized methods.}
	
	\item \textcolor{blue}{This paper designs a general set partitioning strategy that  characterizes sensors by their observable spaces and incorporates the Grassmann distance to achieve robust grouping. This strategy demonstrates strong transferability, enabling the extension of existing neighbor selection–based detection algorithms into a distributed framework.
		 Compared with thenon-partitioned counterpart, the distributed form aligns more naturally with parallel computing architectures, thereby providing an efficient and practical solution for attack detection in large-scale, resource-constrained IoT scenarios.}

	\item \textcolor{blue}{In view of the suitability of submodular optimization theory (SOT) for neighbor-selection problems, this paper extends the SOT-based attack detection scheduling (ADS) algorithm \citep{Suo2024Security} into a two-stage Distributed-ADS (D-ADS) algorithm by incorporating the proposed set partitioning strategy. Both theoretical analysis and simulation results consistently demonstrate that the set partitioning strategy substantially reduces computational cost while maintaining the performance gap between distributed and centralized attack detection within a tolerable range.}

\end{enumerate}

\begin{figure}[htb]
	\centering
	\includegraphics[width=0.8\textwidth]{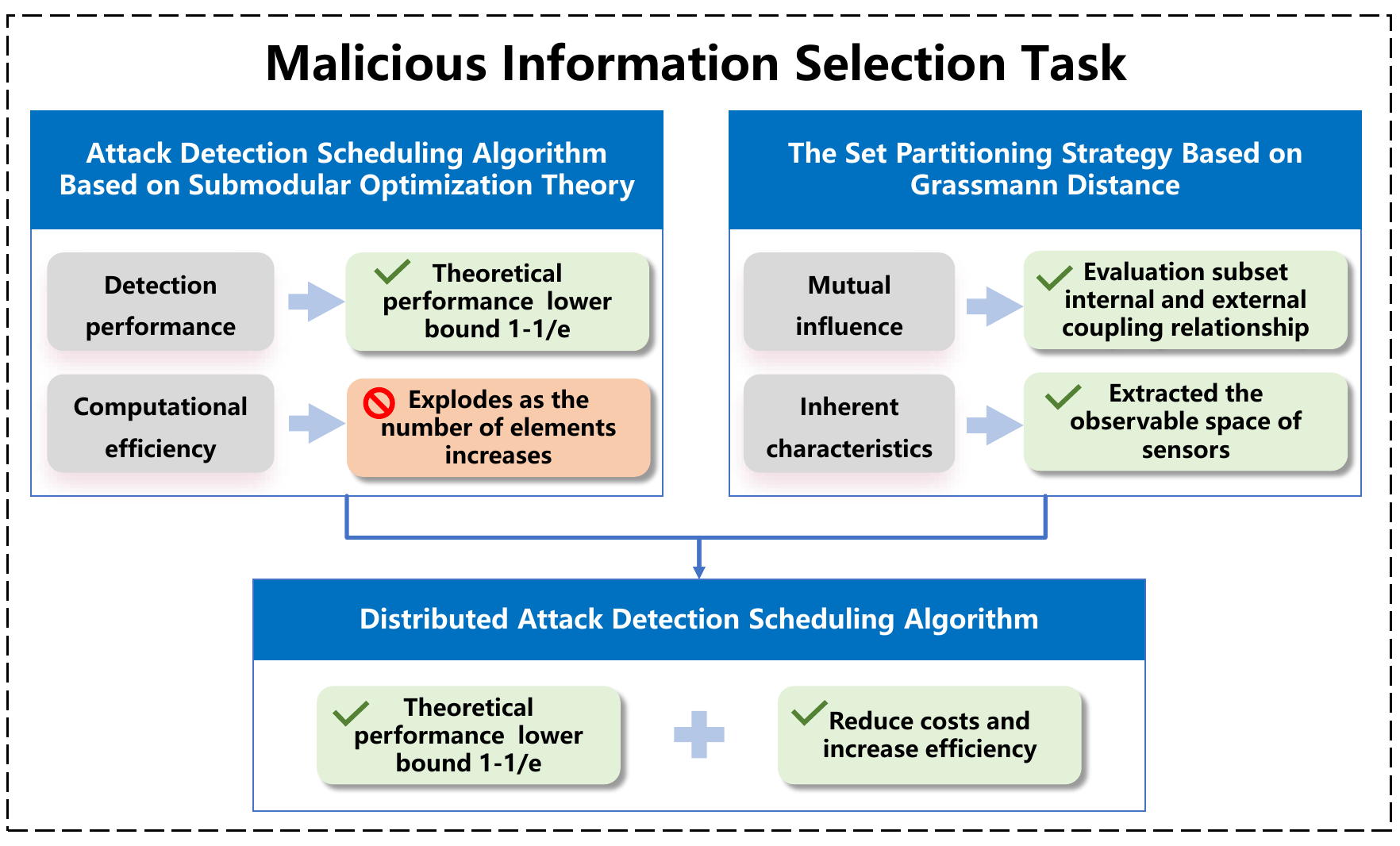}
	\caption{\textcolor{blue}{Relationship diagram between ADS and D-ADS in the malicious information selection task}}
	\label{ADSandDADS}
\end{figure}

\textcolor{blue}{Therefore, the core contribution of this study lies in the proposed set partitioning strategy, which is integrated into a theoretically guaranteed Attack Detection Scheduling (ADS) algorithm. Building upon this foundation, we further extend the ADS algorithm to design a more cost-effective and efficient Distributed-ADS (D-ADS) algorithm. The relationship between ADS and D-ADS is illustrated in Figure \ref{ADSandDADS}.}

This paper is organized as follows: Section II presents the system model and problem formulation, including the description of the sensor network, attacker model, and the malicious information selection problem in dynamic environments. Section III focuses on the analysis of the key factors affecting the performance of distributed algorithms, particularly the mutual influence of benefits between subsets and the design of the set partitioning strategy based on Grassmann distance. Section IV evaluates the performance of the proposed strategies through numerical simulations. Finally, Section V concludes the paper by summarizing the main contributions,  and suggesting potential directions for future research.

\section{Problem Formulation}
\subsection{System Model}
IoT systems usually observe production data and system operation status, which usually changes continuously. Therefore, as a theoretical type of article, the following mathematical model is used to describe this process. Consider a distributed sensor network that monitors the discrete-time linear time-invariant system state $x(k)$, as described in equation (\ref{chapter2:state_equation})
\begin{equation} \label{chapter2:state_equation}
	x\left( k+1 \right) =Ax\left( k \right) +\omega \left( k \right) ,
\end{equation}
where \( x\left( k \right)\in \mathbb{R} ^{n} \) and \( \omega \left( k \right)\in \mathbb{R} ^{n} \) represent the system state and process noise, respectively. 
Moreover, \( \omega \left( k \right) \) follows a Gaussian distribution with zero mean and a positive definite covariance matrix \( Q \), i.e., \( \omega \left( k \right) \sim \mathcal{N} \left( 0,\,Q\right) \).

The network is represented by an undirected graph $\mathcal{G} = (\mathcal{N}, \mathcal{E})$, where $\mathcal{N}$ and $\mathcal{E} \subseteq \mathcal{N} \times \mathcal{N}$ denote the set of sensors and the set of edges, respectively. The neighbor set of sensor $i$ is denoted as $\mathcal{N}_i = \{j \in \mathcal{N} : (i, j) \in \mathcal{E} \}$. Thus, we obtain $\mathcal{N} = \mathcal{N}_1 \cup \mathcal{N}_2 \cup \ldots \cup \mathcal{N}_{|\mathcal{N}|}$. For sensor $i$, where $i \in \mathcal{N}$, its measurement model follows equation (\ref{measurement_equation})
\begin{equation} \label{measurement_equation}
	y_{i}\left( k \right) =C_{i}x\left( k \right) +\nu_{i} \left( k \right) ,
\end{equation}
where \( y_{i}\left( k \right)\in \mathbb{R} ^{m} \) and \( \nu_{i} \left( k \right)\in \mathbb{R} ^{m} \) represent the measurement and measurement noise of sensor \( i \), respectively. The matrix \( A\in\mathbb{R} ^{n\times n} \) in equation (\ref{chapter2:state_equation}) and the matrix \( C_i\in\mathbb{R} ^{m\times n} \) in equation (\ref{measurement_equation}) are both real matrices with appropriate dimensions. 
Similarly, \( \nu_{i} \left( k \right) \) follows a Gaussian distribution with zero mean and a positive definite covariance matrix \( R_{i} \), i.e., \( \nu_{i} \left( k \right) \sim \mathcal{N} \left( 0,R_{i} \right) \). 
In this section, both the process noise \( ||\omega \left( k \right)|| \) and the measurement noise \( ||\nu_{i} \left( k \right)|| \) are upper-bounded by certain positive scalars. 
%For the neighborhood set \( \mathcal{N}_i \) of sensor \( i \), the pair \( ( A, [C_{1}^{\mathrm{T}},...,C_{|\mathcal{N}_i| }^{\mathrm{T}}]^{\mathrm{T}} ) \) is observable.

Furthermore, for the neighbor set $\mathcal{N}_i$ of sensor $i$, the system $( A,\, [C_{1}^{\mathrm{T}},...,C_{|\mathcal{N}_i| }^{\mathrm{T}}]^{\mathrm{T}} )$ is observable. The distributed estimator of sensor $i$ is then given by equation (\ref{distri_estimat})
\begin{equation}
	\label{distri_estimat}
	\hat{x}_i\left( k+1 \right) =A\hat{x}_i\left( k \right) +K_i\left( k \right) \left( y_i\left( k \right) -C_i\hat{x}_i\left( k \right) \right) 
	-\lambda A\sum_{j\in \mathcal{N} _i}{ \left( \hat{x}_i\left( k \right) -\hat{x}_{j}\left( k \right) \right)},
\end{equation}
where \( \hat{x}_i(k) \) is the estimate of the state \( x(k) \) for sensor \( i \), where \( \hat{x}_i(0) = x(0) \). 
Additionally, \( \hat{x}_j(k) \) represents the estimate of sensor \( j \), \( K_i(k) \) is the gain matrix, and \( \lambda \) is the consensus parameter, which takes values within the range \( ( 0, \min(1/|\mathcal{N}_i| )) \). 

\subsection{Attacker Model}
To describe the attacker model in IoT, this paper considers a dynamic attack strategy \citep{Suo2024Security}, At each moment $k$, the attacker selects communication links to launch FDIAs based on the dynamic attack strategy defined in Definition \ref{semi-dynamic attack defn}.
\begin{defn}\label{semi-dynamic attack defn}\textbf{(Dynamic Attack Strategy)}
	\textcolor{blue}{For the sets of compromised sensors at two consecutive moments, \( \mathcal{A}_{k} \) and \( \mathcal{A}_{k-1} \), if they differ, i.e., \( \mathcal{A}_{k} \neq \mathcal{A}_{k-1} \), we refer to this as a dynamic attack strategy. The difference between the two sets can be calculated as}
	\begin{equation}
		\varDelta _k=( \mathcal{A}_{k}\backslash \mathcal{A}_{k-1} ) \cup ( \mathcal {A}_{k-1}\backslash\mathcal{A}_{k} ).
	\end{equation}
	\textcolor{blue}{Thus, over the entire time period \( T \), the total number of changes in compromised sensors is given by \( \varDelta _T=\sum_{k=1}^{T-1}{\varDelta _k} \).}
\end{defn}

At moment $k$, suppose that the estimated state $\hat{x}_j(k)$ of sensor $j$ is compromised by injecting malicious data $z_{ij}(k)$ during its transmission to sensor $i$. The compromised estimation received by sensor $i$, denoted as $\hat{x}_{ij}^{a}(k)$, is given by
\begin{equation}
	\hat{x}_{ij}^{a}(k) = \hat{x}_j(k) + z_{ij}(k).
\end{equation}

In addition, the attacker's strategy satisfies the following Assumption \ref{chapter2:attack_assum}.

\begin{assum}\label{chapter2:attack_assum}
	For the attacker's strategy, the following assumptions are satisfied
	
	\begin{itemize}
		\item At any moment \( k \), the number of compromised sensors \( q_i \) in the neighborhood of sensor \( i \) does not exceed half the size of the neighborhood, i.e., \( q_i \le \lfloor |\mathcal{N}_i|/2 \rfloor \)\citep{yang2021secure,lu2023polynomial}
		\item At any moment \( k \), if the attacker reselects the set of compromised sensors, the selection process is entirely random, and the attack intensity is balanced.
	\end{itemize}
\end{assum}

\subsection{Malicious Information Selection Problem}

Existing literature shows that selecting a set of malicious information from the set $\mathcal{N}_i$ is an NP-hard problem, and the objective function can be converted into a submodular function \citep{Suo2024Security}
%Lemma 1 in literature \cite{Suo2024Security} states that selecting the suspicious sensor set from the set $\mathcal{N}_i$ is an NP-hard problem and that the objective function can be transformed into a submodular function as shown in equation (\ref{f_A})
\begin{equation} \label{f_A}
	%\label{sub_func}
	f_k\left( \mathcal{A}_{i,\,k} \right) =\left\| \varLambda_{i,\,k} \cdot [\mu_{ij}(k)]_{j\in\mathcal{N}_i} \right\|,
\end{equation}
where $\varLambda_{i,\,k}$ is the augmented error
matrix, which is obtained by summarizing the estimated error $||\hat{x}_i( k )- \hat{x}_{ij}^{a}(k) ||$, i.e., $\varLambda_{i,\,k} =\sum^{|\mathcal{N}_i|}_{j=1}{\left(\theta_{|\mathcal{N}_i|}^j\otimes||\hat{x}_i\left( k \right) -\hat{x}_{ij}^{a}\left( k \right)||\right)}$, and $[\mu_{ij}(k)]_{j\in\mathcal{N}_i}$ denotes the augmented matrix of $\mu_{ij}(k)$, where $\mu_{ij}(k)=1$ for $j\in \mathcal{ A}_{i,k}$ and $ \mu_{ij}(k)=0$ for $j\in \mathcal{N}_i\backslash\mathcal{A}_{i,\,k}$.
Consequently, the malicious information selection problem is reformulated as a solvable submodular maximization problem under a cardinality constraint, as presented in Problem \ref{chapter3:prob31}.
\begin{prob}\citep{Suo2024Security}\label{chapter3:prob31}
	% According to Assumption \ref{chapter2:attack_assum}, the number of compromised sensors in the neighborhood set \( \mathcal{N}_i \) of sensor \( i \) does not exceed \( q_i \le \lfloor |\mathcal{N}_i|/2 \rfloor \). 
	The problem of selecting malicious information essentially involves selecting no more than \( q_i \) sensors to maximize the objective function (\ref{f_A}), i.e.,
	\begin{equation} \label{problem1}
		\max_{\mathcal{A}_{i,\,k} \subseteq \mathcal{N}_i} f_k\left( \mathcal{A}_{i,\,k} \right), ~~\text{s.t.} \left| \mathcal{A}_{i,\,k} \right|\leq q_{i},
	\end{equation}
	where \( q_{i} \) is defined in Assumption \ref{chapter2:attack_assum}. 
	%It should be noted that while both equation (\ref{origin_func}) and Equation (\ref{f_A}) aim to select the suspicious sensor set, their function values are not equivalent.
\end{prob}

\section{Main Results}
%It is shown that distributed algorithms have obvious advantages over centralized algorithms in terms of efficiency, but from the perspective of performance, distributed algorithms generally have a certain gap with centralized algorithms. Therefore, this section focuses on the following issues: analyzing the key factors that affect the performance difference between D-ADS and ADS algorithm, and designing effective partitioning strategies to minimize the performance gap.

\subsection{The mutual influence of benefits between subsets}
In this section, the neighbor set of each sensor is  divided into several subsets.
Specifically, the neighbor set $\mathcal{N}_i$ of sensor $i$ is divided into $m_i$ subsets, expressed as $\mathcal{N} _i=\mathcal{N} _{i,\,1}\cup \mathcal{N} _{i,\,2}\cup \cdots \cup \mathcal{N} _{i,\,m_i}$. It should be noted that the parameter $m_i$ is predetermined and has nothing to do with $|\mathcal{N} _i|$.

The method of approximate average partitioning is adopted to ensure that the cardinality of each subset is approximately equal\footnote{At this point, the balance between the number of attacked sensors in each subset and the attack intensity can be obtained intuitively.  Please refer to the Lemma 2 in APPENDIX A.}.  The cardinality $|\mathcal{N} _{i,\,g_i}|$ of the $g_i$-th subset can be expressed as:
\begin{equation}\label{divide_card}
	|\mathcal{N} _{i,\,g_i}|=\lfloor|\mathcal{N}_i|/m_i \rfloor + \mathbb{I}(g_i\le(|\mathcal{N}_i|\  \text{mod}\  m_i|)),
\end{equation}
where $g_i=1:m_i$, and the indicator function $\mathbb{I}(\cdot)$ is defined as follows: when $g_i$ is less than or equal to the remainder of $|\mathcal{N}_i|$ divided by $m_i$, $\mathbb{I}(\cdot)$ is $1$, otherwise $\mathbb{I}(\cdot)$ is $0$. 
% The purpose of this function is to ensure that the entire set is approximately evenly distributed to each subset without any remaining or duplicated elements. 
% In particular, when $|\mathcal{N}_i|$ is divisible by $m_i$, the indicator function $\mathbb{I}(\cdot)$ must be $0$, and the cardinality of each subset is exactly the same. 

% At this point, the balance between the number of attacked sensors in each subset and the attack intensity can be obtained intuitively. And the results from literature \cite{Suo2024Security} indicate that the D-ADS algorithm guarantees a performance lower bound for each subset consistent with that of the ADS algorithm\footnote{The proof process is omitted here. Please refer to the Lemma 2 in APPENDIX A of the full version of this paper\hl{[]}.}.

However, to ensure this performance lower bound, the gain values of all sensors need to be updated after each selection, and existing literature shows that this will bring huge computational complexity \citep{mirzasoleiman2016distributed}. In fact, at the $l$-th selection at moment $k$, only the sensor $j_{g_i^s,\,select}^{l_{g_i^s}}$ in the $g_i^s$-th subset is selected. To reduce the amount of calculation, this section starts with the gain update method and explores the feasibility of only updating  the gains of the remaining elements of the subset where the selected sensor is located.

Based on the inherent characteristics of the observable space of each sensor, this section investigates the mutual influence of gains between sensor subsets and proposes a set partitioning strategy to minimize the mutual influence of gains between subsets.

% First, the  definition of mutual influence is given in Definition \ref{chapter4:defn42}:
% \begin{definition}\label{chapter4:defn42}
	%     \textbf{(mutual influence)} At moment $k$, for any two sensors $j_{g_i}$ and $j_{q_i}$ in any neighbor sensor subsets $\mathcal{N}_{i,\,k,\,g_i}$ and $\mathcal{N}_{i,k,q_i}$ of sensor $i$, the mutual benefit impact between them is defined as the number of dimensions $\Delta\hat{x}_{j_{g_i}}^{re}(k)$ and $\Delta\hat{x}_{j_{q_i}}^{re}(k)$ that are not $0$ simultaneously in all dimensions $\ell=1:n$, denoted as $E(j_{g_i},j_{q_i})=\sum_{\ell=1}^{n}
	%         (\Delta\hat{x}_{j_{g_i}}^{re}(k)[\ell])\land(\Delta\hat{x}_{j_{q_i}}^{re}(k)[\ell]))$.
	% \end{definition}

% Next, the approach to computing the mutual interference between sensor subsets from the perspective of the observable matrix is proposed. 
For the observable discrimination matrix $Q_{j,\,o}$ of neighbor sensor $j \in \mathcal{N}_i$ of sensor $i$, the indicator function for the $\ell$-th row of the $j$-th sensor is defined as a binary indicator function:
% First, the observable discrimination matrix $Q_{j,o}$ for all neighbor sensors $j \in \mathcal{N}_i$ of sensor $i$ is calculated. Then, the indicator function for the $\ell$-th row of the $j$-th sensor is defined as a binary indicator function:
% \begin{equation}\label{indicator_function}
	% 	I_{j,\ell}\left( k \right) =\left\{ \begin{array}{c}
		% 		1, ~~\text{if the} \ \ell \text{-th row of } Q_{j,o} \text{ contains at least one non-zero element,} \\
		% 		0, ~~\text{if the} \ \ell \text{-th row of } Q_{j,o} \text{ contains all zeros.}\\
		% 	\end{array} \right.
	% \end{equation}
\begin{equation}\label{indicator_function}
	I_{j,\,\ell} =
	\begin{cases}
		1, & \exists m \in \{1, \dots, n\}, \ (Q_{j,\,o})_{\ell,\,m} \neq 0, \\
		0, & (Q_{j,\,o})_{\ell,\,:} = 0.
	\end{cases}
\end{equation}

By calculating the indicator function for each row of $Q_{j,\,o}$, a column vector $[I_{j,\,\ell}]_{\ell=1:n} \in \mathbb{R}^{n}$ can be obtained.
Then, the mutual influence is defined in Definition \ref{chapter4:defn42}:
\begin{defn}\label{chapter4:defn42}
	\textbf{(Mutual Influence)} For any two sensors $j_{g_i}$ and $j_{q_i}$ in any neighbor sensor subsets $\mathcal{N}_{i,\,k,\,g_i}$ and $\mathcal{N}_{i,\,k,\,q_i}$ of sensor $i$, the  mutual influence of gains between them is defined as the number of dimensions $[I_{j_{g_i},\,\ell}]_{\ell=1:n}$ and $[I_{j_{q_i},\,\ell}]_{\ell=1:n}$ that are not $0$ simultaneously in all dimensions $\ell=1:n$, denoted as 
	\begin{equation}
		E(j_{g_i},\,j_{q_i})=[I_{j_{g_i},\,\ell}]_{\ell=1:n}\land [I_{j_{q_i},\,\ell}]_{\ell=1:n}.
	\end{equation}
\end{defn}

And the problem of minimizing the mutual influence between subsets is given as shown below:
\begin{prob}\label{chapter3:prob4.1}
	% The problem of minimizing the mutual interference between subsets of benefits is transformed into minimizing the sum of the number of dimensions $\ell=1:n$ where the estimation error vector of sensor $j_{g_i}$ from the $g_i$-th neighbor subset of sensor $i$ and the estimation error vectors of all sensors in the other $m_i-1$ subsets are simultaneously non-zero.
	During the selection of a malicious information, assume that the sensor $j_{g_i}$ from the $g_i$-th neighbor subset of sensor $i$, i.e., $j_{g_i} \in \mathcal{N}_{i,\,k,\,g_i}$, is selected.
	According to equation (\ref{indicator_function}), the $Q_{j,\,o}$ of each sensor $j$ can be transformed into an indicator function vector. Therefore, it is only necessary to ensure that mutual influence between the elements in other subsets $q_i$ and the remaining elements of subset $g_i$ is minimized, that is,
	%Therefore, the objective is to minimize the sum of the number of dimensions $\ell = 1:n$ where the indicator function vector of sensor $j_{g_i}$ and the indicator function vectors of all sensors in the other $m_i - 1$ subsets are simultaneously non-zero, that is,
	\begin{equation}
		\min_{q_i,\,j_{q_i}} \ \ \sum_{q_i=1,\, q_i\neq g_i} ^{m_i}\sum_{j_{q_i}\in\mathcal{N}_{i,\,q_i}}
		E(j_{g_i},\,j_{q_i}).
	\end{equation}
	% \begin{equation}
		%     \min_{q_i,j_{q_i}} \ \ \sum_{q_i=1, q_i\neq g_i} ^{m_i}\sum_{j_{q_i}\in\mathcal{N}_{i,q_i}}
		%     \sum_{\ell=1}^{n}
		%     (I_{j_{g_i,\ell}}(k)\land I_{j_{q_i,\ell}}(k)),
		% \end{equation}
\end{prob}

Based on the observability discrimination matrix, the observable part of each sensor is extracted. As a result, the mutual influence of gains only occurs between sensors whose observable spaces overlap. Therefore, as long as all sensors with overlapping observable spaces are grouped into the same subset, after each malicious information selection, only the gains of the sensors within that subset need to be updated. However, the observable spaces of sensors within the same subset may not be exactly identical. This means that, after each sensor selection, the number of dimensions in the estimation error vector that need to be updated differs across sensors in the subset, which in turn leads to inaccuracies in the allocation ratio vector.

%The observable part of each sensor is extracted based on the observable discrimination matrix, so the mutual influence of gain only occurs between sensors with intersections in the observable space. Therefore, based on the optimal set partitioning strategy, after each malicious information is selected, only the gains of the sensors in the subset need to be updated. However, the observable spaces of the sensors in the same subset may not be exactly the same. This means that after each sensor selection, the number of dimensions that need to be updated in the estimated error vector of each sensor in the subset will be different, resulting in an inaccurate allocation ratio vector $p_{p,\,k}^{(l_p)}$. 
% For example, for a 4-dimensional system state, one sensor can observe 1 and 3 dimensions, while another sensor can only observe 1 dimension. 

\subsection{The  set partitioning strategy based on Grassmann distance}
Based on the previous analysis, this paper considers exploring the partitioning strategy to strike a balance between the mutual influence between subsets and the correlation within subsets.

% Thus, consider introducing a metric that can comprehensively account for both the mutual interference between subsets and the correlation within subsets.
Inspired by the fact that Grassmann distance describes the angular differences in the directions of the vector spanning the subspace, the directional differences between the observable spaces of sensors in the subset are used to evaluate the correlation between sensors.
% Introduce Grassmann distance to measure the minimum angle between two subspaces, i.e., the similarity or dissimilarity between two subspaces.
First, the definition of Grassmann distance is introduced.

\begin{defn}
	\textbf{(Grassmann Distance)} \citep{edelman1998geometry} %Grassmann distance is a method for measuring the similarity between two subspaces. 
	For two subspaces $U$ and $V$, the Grassmann distance $d_G(U,\, V)$ can be calculated by performing singular value decomposition on the bases of these two subspaces, i.e., 
	\begin{equation}\label{chapter4:defn43_equation}
		d_G(U,\,V)= \sqrt{\sum_{i=1}^{m} \theta_i^2},
	\end{equation} 
	where $m$ is the smaller dimension of $U$ and $V$, and $\theta_i$ is the principal angle between $U$ and $V$. Specifically, when the subspaces $U$ and $V$ are each spanned by one-dimensional vectors, $d_G(U,\,V) = \theta$.
\end{defn}

% Next, consider whether the sensor set can be divided based on the Grassmann distance. First, calculate the indicator function column vector $[I_{j,\ell}(k)]_{\ell=1:n}$ for all sensors $j \in \mathcal{N}_i$. Then, calculate the Grassmann distance between every pair of sensor observable spaces and group the completely correlated ones into the same subset. This leads to the following optimization problem \ref{chapter4:prob43}:

Based on the  indicator function column vector $[I_{j,\ \ell}]_{\ell=1:n}$ of all sensors $j \in \mathcal{N}_i$ in equation (\ref{indicator_function}), the Grassmann distance between the observable spaces of each pair of sensors can be obtained. Then, the problem of assigning completely correlated sensors into the same subset can be described as  Problem \ref{chapter4:prob43}:

\begin{prob}\label{chapter4:prob43}
	For the $g_i$-th subset, the Grassmann distance between each pair of sensors $j_{g_i,\, 1}, j_{g_i,\, 2} \in \mathcal{N}_i$ can be obtained. Then, we only need to minimize the Grassmann distances between sensors within each subset $g_i = 1:m_i$ to maximize the intra-subset correlation, i.e.,
	\begin{equation}\label{subset_main_theta}
		\min_{\theta_{j_{g_i,\,1},\, j_{g_i,\,2}}} \, \  \sum_{g_i=1:m_i}   \|[\theta_{j_{g_i,\, 1},\,j_{g_i,\, 2}}]_{j_{g_i,\, 1},j_{g_i,\, 2}\in\mathcal{N}_{i,\, g_i}}\|_2.
	\end{equation}
	%\|\boldsymbol{\theta_{g_i}}\|_2 =
	% \begin{equation}
		%     \max_{\theta_{j_{g_i},j_{q_i}}} \ \  \| \boldsymbol{\theta}\|_2 = \|[\theta_{j_{g_i},j_{q_i}}]_{g_i,q_i \in \{1:m_i\}}\|_2,
		% \end{equation}
	% where $\boldsymbol{\theta} = [\theta_{j_{g_i},j_{q_i}}]_{g_i,q_i \in \{1:m_i\}}$
\end{prob}

For any two sensors $j_1$ and $j_2$ in the set $\mathcal{N}_i$, the cosine of the angle between their indicator function column vectors can be calculated using the vector dot product formula, i.e., 
\begin{equation}
	\label{cos_value}
	\cos(\theta_{j_1,\, j_2}) = \frac{[I_{j_1,\, \ell}]_{\ell=1:n} \cdot [I_{j_2,\, \ell}]_{\ell=1:n}}{\|[I_{j_1,\, \ell}]_{\ell=1:n}\| \cdot \|[I_{j_2,\, \ell}]_{\ell=1:n}\|}.
\end{equation}

The Grassmann distance between the observable spaces of sensors \( j_1 \) and \( j_2 \) can be determined by  
\begin{equation}\label{Grassmann_distance}
	d_G([I_{j_1,\, \ell}]_{\ell=1:n},[I_{j_2,\, \ell}]_{\ell=1:n})= \arccos(\cos(\theta_{j_1,\, j_2})),
\end{equation}  
where the result is given in radians.
\begin{rem}
	It should be noted that equation (\ref{indicator_function}) has already converted rows containing nonzero elements into binary values, either \( 1 \) or \( 0 \). As a result, the Grassmann distance calculated between the indicator function column vectors only has two possible outcomes: either the vectors are perfectly aligned (\(\theta_{j_1,\,j_2}=0\)) or they are orthogonal (\(\theta_{j_1,\,j_2}=\pi/2\)). Ideally, for any \( g_i \)-th subset, the result of equation (\ref{subset_main_theta}) should be zero. 
\end{rem}

Based on the above analysis, the partitioning of the set $\mathcal{N}_i$ aims to find a compromise solution for two objectives: maximizing the intra-subset correlation and minimizing the mutual influence between subsets, and the correlation within the subset has a greater priority.

\begin{algorithm}[htb]  
	%\setstretch{1.3}
	\renewcommand{\thealgorithm}{3.1}
	\caption{The Set Partitioning Strategy Based on Grassmann Distance}  
	\label{alg:3-3}  
	\begin{algorithmic}[1]  
		\Require  The observable discrimination matrix $Q_{j,\, o}$ for all sensors $j \in \mathcal{N}_i$.
		\Ensure  The partitioning result of the sensor set $\mathcal{N}_i$.
		\State Initialize $\mathcal{N}_{i,\, g_i}$, with $g_i=1:m_i$, and set the initial value of $m_i$ to 1, which will increase dynamically.
		\For{$j$ in $\mathcal{N}_i$}
		\For{$\ell=1:n$}
		\State Calculate the indicator function $I_{j,\, \ell}$ for the $\ell$-th row of $Q_{j,\, o}$ based on equation (\ref{indicator_function}).
		\EndFor
		\State Obtain the indicator function column vector $[I_{j,\, \ell}(k)]_{\ell=1:n} \in \mathbb{R}^{n}$.
		\EndFor
		
		\For{$idx_1 = 1:|\mathcal{N}_i|$}
		\State found group = 0.
		\For{$g_i=1:m_i$}
		\State $idx_2 = \mathcal{N}_{i,\, g_i}\{1\}$. $\%$ Take the first element from the set $g_i$.
		% \EndFor
		% \For{$idx_2 = idx_1+1:|\mathcal{N}_i|$}
		\State Calculate the cosine value of the angle between the indicator function column vectors of sensors $j_{idx_1}$ and $j_{idx_2}$, which span a subspace, using equation (\ref{cos_value}).
		\State Calculate the Grassmann distance between the two subspaces using equation (\ref{Grassmann_distance}).
		\If{$d_G([I_{j_{idx_1},\, \ell}]_{\ell=1:n}, [I_{j_{idx_2},\,\ell}]_{\ell=1:n}) == 0$}
		\State $\mathcal{N}_{i,\, g_i} = \mathcal{N}_{i,\, g_i} \cup \{j_{idx_2}\}$.
		\State found group = 1.
		\State break.
		% Sensors $j_{idx_1}$ and $j_{idx_2}$ belong to the same subset.
		% \State $m_i$ 
		% \State Sensors $j_{idx_1}$ and $j_{idx_2}$ do not belong to the same subset.
		\EndIf
		\EndFor
		\If{found group == 0}   $\%$ Create a new subset.
		\State $m_i = m_i + 1$.
		\State $\mathcal{N}_{i,\, m_i} = \mathcal{N}_{i,\, m_i} \cup \{j_{idx_2}\}$.
		\EndIf
		\EndFor
		
		\State \Return{The $g_i$-th sensor subset $\mathcal{N}_{i,\, g_i}$,  $g_i=1:m_i$.}
	\end{algorithmic}  
\end{algorithm}

To reduce the computational cost incurred by calculating the Grassmann distance in step 13 of Algorithm \ref{alg:3-3}, the improved algorithm is proposed in Algorithm \ref{alg:3-4}. First, we present the following Lemma.
\begin{lem}\label{chapter3:lem46}
	A necessary but not sufficient condition for maximizing the intra-subset correlation is that all sensors within the subset have the same observable space dimension.
\end{lem}

\begin{pf}
	%First, the following metric can be used to evaluate the degree of internal correlation within a subset.
	
	For any pair of two elements \( j_{g_i,\, 1} \) and \( j_{g_i,\, 2} \) within the \( g_i \)-th subset of sensor \( i \), 
	% their correlation is determined by comparing their indicator function vectors \( [I_{j_{g_{i},1},\ell}]_{\ell=1:n} \) and \( [I_{j_{g_{i},2},\ell}]_{\ell=1:n} \) to check if they are identical. Consequently, 
	the intra-subset correlation of the \( g_i \)-th subset is defined as
	%\begin{small}
	\begin{equation} \label{chapter3:equ4.19}
		\frac{\sum_{(j_{g_{i},\, 1}, j_{g_{i},\ 2}) \in \mathcal{N}_{i,\, g_i} \times \mathcal{N}_{i,\,g_i}}  \textbf{1}([I_{j_{g_{i},\, 1},\, \ell}]_{\ell=1:n} =
			[I_{j_{g_{i},\, 2},\, \ell}]_{\ell=1:n} )   
		}{|\mathcal{N}_{i,\, g_i}|\cdot|\mathcal{N}_{i,\, g_i}|}.
	\end{equation}
	%\end{small}
	
	According to equation (\ref{chapter3:equ4.19}), when the indicator function vectors of two sensors are identical, they are fully correlated. In this case, their observable space dimensions must be the same. However, the converse does not necessarily hold.  
	This complete the proof.
\end{pf}

Inspired by Lemma \ref{chapter3:lem46}, we first partition all sensors with the same dimension \( Q_{j,\,o} \) into the same initial set. Then, we only need to apply Algorithm \ref{alg:3-3} separately in each set to partition the subsets. Theoretically, this strategy greatly reduces the computational cost required to calculate the Grassmann distance. The details are given in Algorithm \ref{alg:3-4}.
\begin{algorithm}[htb]  
	%\setstretch{1.3}
	\renewcommand{\thealgorithm}{3.2}
	\caption{The improved sensor set partitioning strategy based on Grassmann distance}  
	\label{alg:3-4}  
	\begin{algorithmic}[1]  
		\Require  The Cell array $Q_{total,\, o}$ consists of the observable matrices $Q_{j,\, o}$ of all sensors $j\in\mathcal{N}_i$.
		\Ensure  Set partitioning results of sensor set $\mathcal{N}_i$.
		% \For{$j\in\mathcal{N}_i$}
		% \State 计算$rank(Q_{j,o})$。
		% \EndFor
		% \State 根据各个传感器的能观测判别矩阵维数初步分组。

		% 初始化一个空的 Map 对象，键是秩，值是矩阵的 cell 数组
		\State Initialize an empty Map object $rankGroups$.
		% $\%$ rankGroups的键是秩，值是矩阵的Cell数组
		
		% 遍历输入的矩阵 cell 数组
		\For{$j = 1:|\mathcal{N}_i|$}
		\State Obtain the observable matrix of sensor $j$ by $Q_{j,\,o}=Q_{total,\,o}\{j\}$.
		\State Calculate the rank $rank(Q_{j,\, o})$ of each observable matrix $Q_{j,\,o}$.
		
		% 检查当前秩是否已存在于 Map 中
		\If{$isKey(rankGroups, rank(Q_{j,\, o}))$}
		% 如果已存在，添加当前矩阵到对应的 cell 数组
		\State $rankGroups(rank(Q_{j,o})) = [rankGroups(rank(Q_{j,\, o}))\  \{Q_{j,\, o}\}]$.
		\Else
		% 如果不存在，初始化一个新的 cell 数组，并添加当前矩阵
		\State Create a key-value pair $rankGroups(rank(Q_{j,\ o})) = {Q_{j,\, o}}$.
		\EndIf
		\EndFor
		
		\State $keys = rankGroups.keys$.
		\For{$r = 1:length(keys)$}
		\State Output all observable matrices $Q_{j,\, o}\in rankGroups(keys\{r\}))$ to Algorithm \ref{alg:3-3}, and obtain the set partitioning result.
		\EndFor

		\State \Return{The $g_i$-th sensor subset $\mathcal{N}_{i,\, g_i}$, $g_i=1:m_i$.}
	\end{algorithmic}  
\end{algorithm}

On this basis, the ADS algorithm in the literature \cite{Suo2024Security} can be deployed on each subset. At this time, the ADS algorithm becomes a distributed case, that is, the D-ADS algorithm. Theoretical results show that the D-ADS algorithm guarantees the same performance lower bound for each subset as the ADS algorithm. The proof process is omitted here. Please refer to the APPENDIX B.

\begin{rem}(\textbf{Computational Cost})
	Given that the indicator function column vectors have the same dimension, the computational cost is proportional to the number of Grassmann distance calculations. In Algorithm \ref{alg:3-3}, the number of Grassmann distance calculations is \( \binom{|\mathcal{N}_i|}{2} \). Suppose the sensor set \( \mathcal{N}_i \) can be divided into \( m_i^{\prime} \) subsets based on dimension, then each subset contains approximately \( |\mathcal{N}_i|/m_i^{\prime} \) sensors. In the improved algorithm, the required number of Grassmann distance calculations is
	$ m_i\cdot\binom{|\mathcal{N}_i|/m_i^{\prime}}{2}$.
	Comparing the two, the reduction in the number of Grassmann distance calculations is:
	\begin{equation}
		\Delta C = \frac{|\mathcal{N}_i|^2(m_i^{\prime}-1)+|\mathcal{N}_i|}{2m_i^{\prime}}.
	\end{equation}
	Notably, when \( m_i^{\prime} > 1 \), this reduction is significant, implying that the improved algorithm effectively reduces the computational cost.
\end{rem}

\begin{rem}\label{chapter3:remark4.1}
	The proposed Algorithm \ref{alg:3-4} can ensure that the sensor observable space of each subset is consistent, but based on the D-ADS Algorithm, the theoretical performance of each subset is guaranteed only when the number of sensors in each subset is equal. Therefore, in the offline pre-setting stage, we need to ensure that the number of sensors with different observation spaces is equal, which is feasible.
\end{rem}

% Theorem \ref{thm48} will prove that, based on the proposed set partitioning strategy, only the gains of the remaining sensors within the subset where the selected sensor is located need to be updated, without updating the gains of all remaining sensors.
\subsection{Theoretical performance}
Theorem \ref{thm48} will prove that based on the proposed set partitioning strategy, although the computational cost of updating the gain is reduced, the impact on the sensor selection performance is limited.
\begin{thm}\label{thm48}
	For the neighbour set $\mathcal{N}_i$ of sensor $i$, the set $\mathcal{N}_i$ is divided by Algorithm \ref{alg:3-4}. At the $l-1$-th selection, suppose that an element $j_s^{(l-1)}$ is selected from one of the subset, and only the gains of the remaining sensors in the subset where the element is located are updated. Then, at the $l$-th selection, the distribution ratio vector of the sensor selection of any $g_i$-th subset $\mathcal{N}_{i,\,k,\,g_i}$ is accurate, or the error is tolerable.
\end{thm}

\begin{pf}
	According to the aforementioned analysis, there are $3$ types of relationships between two sensors:  Completely correlated, 
	Partially correlated,
	Completely uncorrelated.
	% In the case of full correlation, the two sensors have exactly the same observation space; in the case of partial correlation, the observation spaces of the two sensors have an intersection, but are not exactly the same; and in the case of complete uncorrelation, the observation spaces of the two sensors do not have any intersection. 
	Based on the set partitioning strategy shown in algorithm \ref{alg:3-4}, the sensors within each subset are completely correlated, while these sensors are partially correlated or completely uncorrelated with the sensors in other subsets.
	
	% Algorithm \ref{alg:3-4} ensures that the sensors within each subset are fully correlated, while these sensors are partially correlated or completely uncorrelated with the sensors of other subsets.
	
	The distribution ratio vector error under $3$ types of  relationships are analyzed as follows:

	For the case where the $l-1$-th selected sensor is completely uncorrelated with the $l$-th selected sensor, that is, there is no intersection in the observation spaces of the two sensors, the gain does not need to be updated at this time, and the distribution ratio vector error is $0$.
	
	% For the case where the $l-1$-th and $l$-th selected sensors are completely uncorrelated, that is, there does not exist mutual influence between the two sensors. At this time, the gain does not need to be updated, and the distribution ratio vector error is $0$. 
	
	% For the case where the $l-1$-th and $l$-th selected sensors are completely correlated, at this time, the sensor gain is updated. Therefore, the distribution ratio vector error is also $0$.
	
	For the case where the $l-1$-th selected sensor is completely correlated with the $l$-th selected sensor, that is, the observation spaces of the two sensors are exactly the same, the sensor gains are updated. Therefore, the distribution ratio vector error is also $0$.
	
	However, for the case where the $l-1$-th selected sensor is partially correlated with the $l$-th selected sensor, that is, the observation spaces of the two sensors intersect but are not exactly the same, so the gains of the two sensors have mutual influence. However, to reduce the computational cost, after the $l-1$th selection, the gain of the $l$th selected sensor is not updated, so there must be an error in the distribution ratio vector. The following proves that the distribution ratio vector error is bounded in this case.
	
	% However, for the case where the $l-1$-th and $l$-th selected sensors are partially correlated, there exists a mutual influence between the two sensors. Therefore, the gain also has a marginal benefit effect. However, to reduce the amount of calculation, the gain is not updated in D-ADS algorithm. Therefore, there must be an error in the distribution ratio vector. The following proves that the distribution ratio vector error in this case is bounded. 
	
	For the $l-1$-th selected sensor $j_s^{(l-1)}$ and the $l$-th selected sensor $j_s^{(l)}$, the indicator function vectors are partially correlated. Considering the effect of diminishing marginal returns, the gain is most affected when the sensor is initially selected, and gradually decreases for subsequent selections. Therefore, the distribution ratio vector error is tolerable as long as the effect of the $l-1=1$-th selected sensor $j_s^{(1)}$ on the distribution ratio vector error of the $l=2$-th selected sensor $j_s^{(2)}$ is bounded.
	
	%     For the $l-1$-th selected sensor $j_s^{(l-1)}$ and the $l$-th selected sensor $j_s^{(l)}$, the indicator function vectors are partially correlated. Considering the impact of diminishing marginal benefits, the gain is most affected when the sensor is initially selected.
	% Therefore, as long as the $l-1=1$-th selected sensor $j_s^{(1)}$'s impact on the distribution ratio vector error of the $l=2$-th selected sensor $j_s^{(2)}$ is bounded, then the distribution ratio vector error is tolerable. 

	Assuming that the diminishing marginal benefit of selecting sensor $j_s^{(1)}$ on the gain of $j_s^{(2)}$ is ignored, the gain $G_{k,	\, j_s^{(2)}}^{(2)}$ of sensor $j_s^{(2)}$ is calculated as $G_{k,\, j_s^{(2)}}^{(2)}=f_k(\{j_s^{(2)}\})$. However, if the influence of selecting element $j_s^{(1)}$ on the gain of element $j_s^{(2)}$ is considered, the gain $G_{k,\, j_s^{(2)}}^{(2)}$ of sensor $j_s^{(2)}$ is calculated as $G_{k,\, j_s^{(2)}}^{(2)}=f_k(\mathcal{A}_{i,k}^{(1)}\cup\{j_s^{(2)}\})-f_k(\mathcal{A}_{i,k}^{(1)})$. The absolute value of the gain error in the two cases is $\Delta_{j_s^{(1)},\, j_s^{(2)}}$, which actually indicates the change in the gain of $j_s^{(2)}$ caused by the influence of $j_s^{(1)}$ on the gain.
	Therefore, in the two cases, the distribution ratio vectors of the subset where the element $j^{(2)}$ belongs are respectively as shown below:
	%\vspace*{-10pt}%
	
	%\begin{widetext}
		%\leftsep
		{
		\begin{equation}\label{equ424}
			p^{\prime}=\frac{[f_k(\{j^{(2)}\}]_{j^{(2)}\in\mathcal{N}_{i,\,k,\,g_i}}}{\sum_{j^{(2)}\in\mathcal{N}_{i,\,k,\,g_i}}f_k(\{j^{(2)}\})},
		\end{equation}   
		\begin{equation}\label{equ425}
		 p^{\prime\prime}=\frac{[f_k(\{j^{(2)}\})+\Delta_{j_s^{(1)},j^{(2)}}]_{j^{(2)}\in\mathcal{N}_{i,\,k,\,g_i}}}{\sum_{j^{(2)}\in\mathcal{N}_{i,\,k,\,g_i}}(f_k(\{j^{(2)}\})+\Delta_{j_s^{(1)},\,j^{(2)}})}.
		\end{equation}  } 
		%\leftsep
		%\rightsep
		
		%\begin{multicols}{2}
		Therefore, the distribution ratio error $|p^{\prime}(j_s^{(2)})-p^{\prime\prime}(j_s^{(2)})|$ of any element $j_s^{(2)}$ in the set $\mathcal{N}_{i,\,k,\,g_i}$ is
		%\end{multicols}

		%\leftsep
%		\textcolor{blue}{
%		\begin{eqnarray}\label{distribution_error}
%			&&|p^{\prime}(j_s^{(2)})-p^{\prime\prime}(j_s^{(2)})|\nonumber\\
%			&=&\frac{|f_k(\{j_s^{(2)}\}\cdot\sum_{j^{(2)}\in\mathcal{N}_{i\,,k,g_i}}\Delta_{j_s^{(1)},j^{(2)}}-\sum_{j^{(2)}\in\mathcal{N}_{i,\,k,\,g_i}}f_k(\{j^{(2)}\})\cdot\Delta_{j_s^{(1)},j_s^{(2)}}|}{(\sum_{j^{(2)}\in\mathcal{N}_{i,\,k,\,g_i}}f_k(\{j^{(2)}\}))\cdot(\sum_{j^{(2)}\in\mathcal{N}_{i,\,k,\,g_i}}(f_k(\{j^{(2)}\})+\Delta_{j_s^{(1)},j^{(2)}}))}\nonumber\\
%			&\le&\frac{\max_{j^{(2)}} \Delta_{j_s^{(1)},j^{(2)}}\cdot|f_k(\{j_s^{(2)}\}\cdot|\mathcal{N}_{i,\,k,\,g_i}|-\sum_{j^{(2)}\in\mathcal{N}_{i,\,k,\,g_i}}f_k(\{j^{(2)}\})|}{(\sum_{j^{(2)}\in\mathcal{N}_{i,\,k,\,g_i}}f_k(\{j^{(2)}\}))\cdot(\sum_{j^{(2)}\in\mathcal{N}_{i,\,k,\,g_i}}(f_k(\{j^{(2)}\})+\Delta_{j_s^{(1)},j^{(2)}}))},
%		\end{eqnarray} }
	{
	\begin{eqnarray}\label{distribution_error}
		&& \bigl| p^{\prime}(j_s^{(2)}) - p^{\prime\prime}(j_s^{(2)}) \bigr| \nonumber\\
		&=& \frac{ \bigl| f_k(\{ j_s^{(2)} \}) \cdot \sum_{j^{(2)} \in \mathcal{N}_{i,\,k,\,g_i}} \Delta_{\,j_s^{(1)},\,j^{(2)}} 
			\;-\; \sum_{j^{(2)} \in \mathcal{N}_{i,\,k,\,g_i}} f_k(\{ j^{(2)} \}) \cdot \Delta_{\,j_s^{(1)},\,j_s^{(2)}} \bigr|}
		{ \bigl( \sum_{j^{(2)} \in \mathcal{N}_{i,\,k,\,g_i}} f_k(\{ j^{(2)} \}) \bigr)\,\cdot\,
			\bigl( \sum_{j^{(2)} \in \mathcal{N}_{i,\,k,\,g_i}} ( f_k(\{ j^{(2)} \}) + \Delta_{\,j_s^{(1)},\,j^{(2)}} ) \bigr)} \nonumber\\
		&\le& \frac{ \max_{j^{(2)}} \Delta_{\,j_s^{(1)},\,j^{(2)}} \,\cdot\, 
			\bigl| f_k(\{ j_s^{(2)} \}) \cdot |\mathcal{N}_{i,\,k,\,g_i}| 
			\;-\; \sum_{j^{(2)} \in \mathcal{N}_{i,\,k,\,g_i}} f_k(\{ j^{(2)} \}) \bigr| }
		{ \bigl( \sum_{j^{(2)} \in \mathcal{N}_{i,\,k,\,g_i}} f_k(\{ j^{(2)} \}) \bigr)\,\cdot\,
			\bigl( \sum_{j^{(2)} \in \mathcal{N}_{i,\,k,\,g_i}} ( f_k(\{ j^{(2)} \}) + \Delta_{\,j_s^{(1)},\,j^{(2)}} ) \bigr)} ,
	\end{eqnarray}}
		%\leftsep
		%\rightsep
		~\\where the  inequality holds because of $\Delta _{j_s^{(1)},\, j_s^{(2)}}\le\max_{j^{(2)}} \Delta_{j_s^{(1)},\,j^{(2)}}=\Delta_{j_s^{(1)},\,j_{max}^{(2)}}$.
		
		%\begin{multicols}{2}
		First, consider the special case where there is \emph{no attack}. At this time, the gain of each sensor is only affected by noise. Therefore, from the perspective of the entire time period, the expectation of the gain $f_k(\{j^{(2)}\})$ of each element is theoretically $0$, that is, $|f_k(\{j_s^{(2)}\}\cdot|\mathcal{N}_{i,\,k,\,g_i}|-\sum_{j^{(2)}\in\mathcal{N}_{i,\,k,\,g_i}}f_k(\{j^{(2)}\})|$ is also close to $0$, which means that the increase of element $j_s^{(1)}$ leads to a negligible change in the gain of element $j^{(2)}$.
		
		Furthermore, consider the case where there \emph{exists attack}. Assume that $q_{i,\,g_i}$ sensors in the $g_i$-th subset are attacked, while the remaining $|\mathcal{N}_{i,\,k,\,g_i}|-q_{i,\,g_i}$ sensors are normal.
				%Divide the equation (\ref{distribution_error}) into two parts:
				Then, the equation (\ref{distribution_error}) is divided into two parts:
	 $|f_k(\{j_s^{(2)}\}\cdot|\mathcal{N}_{i,\,k,\,g_i}|-\sum_{j^{(2)}\in\mathcal{N}_{i,\,k,\,g_i}}f_k(\{j^{(2)}\})|$ and $\Delta_{j_s^{(1)},\,j_{max} ^{(2)}}/\left((\sum_{j^{(2)}\in\mathcal{N}_{i,\,k,\,g_i}}f_k(\{j^{(2)}\}))\cdot(\sum_{j^{(2)}\in\mathcal{N}_{i,\,k,\,g_i}}(f_k(\{j^{(2)}\})+\Delta_{j_s^{(1)},j^{(2)}}))\right)$.
		
		For the former, define a $0-1$ binary parameter $\rho$, whose values $0$ and $1$ represent the states of sensor $j_s^{(2)}$ is under attack (denoted as $j_s^{(2),\,u}$) and without attack (denoted as $j_s^{(2),\,w}$), respectively. At this time, the following derivation is obtained
		%\end{multicols}
		%\leftsep
		\begin{eqnarray}\label{chapter3:proof_fangsuo}
			&&|f_k(\{j_s^{(2)}\}\cdot|\mathcal{N}_{i,\,k,\,g_i}|-\sum_{j^{(2)}\in\mathcal{N}_{i,\,k,\,g_i}}f_k(\{j^{(2)}\})|\nonumber\\
			&=&|f_k(\{j_s^{(2)}\}\cdot|\mathcal{N}_{i,\,k,\,g_i}|-\sum_{j^{(2),w}\in\mathcal{N}_{i,\,k,\,g_i}\backslash I_{i,\,k,\,g_i}}f_k(\{j^{(2),\,w}\})-\sum_{j^{(2),\,u}\in I_{i,\,k,\,g_i}}f_k(\{j^{(2),\,u}\})  |\nonumber\\
			&\le&\rho \cdot \left( |\mathcal{N}_{i,\,k,\,g_i}|-q_{i,\,g_i} \right) |\min_{j^{(2),\,w}\in\mathcal{N}_{i,\,k,\,g_i}\backslash I_{i,\,k,\,g_i}} f_k(\{j^{(2),\,w}\}) -\max_{j^{(2),\,u}\in I_{i,\,k,\,g_i}}f_k(\{j^{(2),u}\}) |\nonumber\\
			&&+(1-\rho)\cdot q_{i,\,g_i} \cdot|\max_{j^{(2),u}\in I_{i,\,k,\,g_i}} f_k(\{j^{(2),\,u}\}) -\min_{j^{(2),\,w}\in\mathcal{N}_{i,\,k,\,g_i}\backslash I_{i,\,k,\,g_i}} f_k(\{j^{(2),\,w}\}) |\nonumber\\
			&=& \max\{|\mathcal{N}_{i,\,k,\,g_i}|-q_{i,\,g_i} ,q_{i,\,g_i}\} |f_k(\{j_{max}^{(2),\,u}\}) -f_k(\{j_{min}^{(2),\,w}\}) |,
		\end{eqnarray}
		%\leftsep
		%\rightsep
		%\begin{multicols}{2}
		where $I_{i,\,k,\,g_i}$ in the first equation represents the set of all attacked sensors in the set $N_{i,\,k,\,g_i}$, and the first inequality holds because the maximum error that the selected sensor may bring is the maximum gain in the attacked sensor set and the minimum gain in the normal sensor set. In the second equation, $j_{max}^{(2),\,u}$ and $j_{min}^{(2),\,w}$ are equal to $\max_{j^{(2),\,u}\in I_{i,\,k,\,g_i}} f_k(\{j^{(2),\,u}\})$ and $\min_{j^{(2),\,w}\in\mathcal{N}_{i,\,k,\,g_i}\backslash I_{i,\,k,\,g_i}} f_k(\{j^{(2),\,w}\})$, respectively.
		%\end{multicols}
		%\leftsep
		
		In addition, the parameter $\mu _{ij}^{(u)}(k)$ ($\mu _{ij}^{(w)}(k)$) is defined as the estimation error when the communication link between sensor $j$ and $i$ is under attack (without attack), which only depends on the network effect and the measurement noise and process noise of sensors $i$ and $j$. By taking the expectation of equation (\ref{chapter3:proof_fangsuo}) and omitting the subscripts $max$ and $min$, we can get
		\begin{eqnarray}
			%&&\mathbb{E}[|\sum_{A\in \mathcal{N}_{i,\,k,\,g_i}}{g_A}(S\cup \{B\})-g_A(S\cup \{B\})\cdot |\mathcal{N}_{i,\,k,\,g_i}||]\nonumber\\
			&& \mathbb{E}\left[\max\{|\mathcal{N}_{i,\,k,\,g_i}|-q_{i,\,g_i} ,q_{i,\,g_i}\}\cdot |f_k(\{j^{(2),\,u}\}) -f_k(\{j^{(2),\,w}\}) |\right] \nonumber\\
			% &\le& \mathbb{E}\left[max\{|\mathcal{N}_{i,\,k,\,g_i}|-q_{i,g_i} ,q_{i,g_i}\} (z_{ij}(k)+\mu_{ij}(k))\right] \nonumber\\
			&\le& \mathbb{E}\left[\max\{|\mathcal{N}_{i,\,k,\,g_i}|-q_{i,\,g_i} ,q_{i,\,g_i}\}\cdot \frac{|f_k(\{j^{(2),\,u}\})^2 -f_k(\{j^{(2),\,w}\})^2 |}{|f_k(\{j^{(2),\,u}\}) +f_k(\{j^{(2),\,w}\}) |}\right] \nonumber\\
			&=&\mathbb{E}\left[\max\{|\mathcal{N}_{i,\,k,\,g_i}|-q_{i,\,g_i} ,q_{i,\,g_i}\}\cdot \frac{(z_{i,\,j^{(2),u}}+\mu_{i,\,j^{(2),\,u}}^{u})^2-(\mu_{i,\,j^{(2),\,w}}^{w})^2}{|z_{i,\,j^{(2),\,u}}+\mu_{i,\,j^{(2),\,u}}^{u}+\mu_{i,\,j^{(2),w}}^{w}|}\right] \nonumber\\
			&\le&   \mathbb{E}\left[\max\{|\mathcal{N}_{i,\,k,\,g_i}|-q_{i,\,g_i} ,\,q_{i,\,g_i}\}\cdot \frac{(z_{i,\,j^{(2),\,u}})^2+2z_{i,\,2}\mu_{i}^{u}+(\mu_{i}^{u})^2-(\mu_{i}^{w})^2}{|z_{i,\,j^{(2),\,u}}+\mu_{i,\,j^{(2),\,u}}^{u}+\mu_{i,\,j^{(2),\,w}}^{w}|}\right]  \nonumber\\
			&\le& \max \left\{\lfloor\frac{|\mathcal{N}_i|-q_i}{m_i} \rfloor + \mathbb{I}\,(g_i\le(|\mathcal{N}_i|\, mod\, m_i|)) ,\,\frac{q_i}{m_i}\right\} \cdot \frac{(\sum_{l=1}^{q_i}\phi_{i,\,q_i}(k)/q_i+2||\mu_{i}(k)||^2_{\infty})}{\bar{\mu_{i}}(k)} ,
		\end{eqnarray}
		%\leftsep
		%\rightsep
		%\begin{multicols}{2}
		where the first inequality is obtained by multiplying both the numerator and the denominator by $|f_k(\{j^{(2),\,u}\}) +f_k(\{j^{(2),\,w}\}) |$. The first equality holds because the values of $f_k(\{j^{(2),\,u}\})$ and $f_k(\{j^{(2),\,w}\})$ are essentially related to the attack signal $z_{i,\,j^{(2),\,u}}$ and the estimation error $\mu_{i,\,j^{(2),\,u}}^u$ ($\mu_{i,\,j^{(2),\,w}}^w$). The third inequality holds because $z_{i,\,j^{(2),\,u}}(k)$ and $\mu_{i,\,j^{(2),\,u}}^u(k)$ are completely independent over the entire time period, and the expectation of the attack signal $z_{i,\,j^{(2),\,u}}^2$ on any sensor $j^{(2),\,u}$ is approximately $1/q_i$, which is the sum of the average malicious perturbation power of all attacks, expressed as $\mathbb{E}[z_{i,\,j^{(2),\,u}}(k)^2]=\sum_{l=1}^{q_i}\phi_{i,\,q_i}(k)/q_i$. In addition, the estimated error $\mu _{ij}\left( k \right)$ satisfies $||\mu _i\left( k \right) ||_{\infty}=\max _{j\in \mathcal {N} _i}\left\{ ||\mu _{ij}\left( k \right) ||_{\infty}\right\} $ and the estimation error $\mu _{ij}\left( k \right)$ satisfies $|z_{i,\,j^{(2),\,u}}(k)|\ge\left\|\mu_i(k) \right\| _{\infty}\ge |\sum_{j\in \mathcal{N}_i}{\mu_{ij}(k)} |/|\mathcal{N}_i|=\bar{\mu_i}$.
		%\end{multicols}
		%\leftsep
		
		For the latter of equation (\ref{distribution_error}), dividing both the numerator and denominator by $\Delta_{j_s^{(1)},j_{max}^{(2)}}$, we get
		\begin{equation}\label{equ427}
			1/\left((\sum_{j^{(2)}\in\mathcal{N}_{i,\,k,\,g_i}}f_k(\{j^{(2)}\}))\cdot\left(\frac{\sum_{j^{(2)}\in\mathcal{N}_{i,\,k,\,g_i}}(f_k(\{j^{(2)}\})+\Delta_{j_s^{(1)},j^{(2)}})}{\Delta_{j_s^{(1)},j_{max}^{(2)}}}\right)\right).
		\end{equation}
		Since $\sum_{j^{(2)}\in\mathcal{N}_{i,\,k,\,g_i}}(f_k(\{j^{(2)}\})+\Delta_{j_s^{(1)},\,j^{(2)}})\gg f_k(\{j^{(2)}\})+\Delta_{j_s^{(1)},\,j^{(2)}}> \Delta_{j_s^{(1)},\,j_{max}^{(2)}}$ always holds, equation (\ref{equ427}) is an infinitesimal positive number.

		%\leftsep
	%\end{widetext}

	Considering the two parts of equation (\ref{distribution_error}), ideally, $|p^{\prime}(j_s^{(2)})-p^{\prime\prime}(j_s^{(2)})|\rightarrow 0$, because a bounded positive number multiplied by an infinitesimal number tends to $0$. In practice, we only need to ensure $|p^{\prime}(j_s^{(2)})-p^{\prime\prime}(j_s^{(2)})|\le \epsilon$, which is obviously achievable. This also means that based on the proposed set partitioning strategy, only the gains of the remaining sensors in the subset where the selected sensor is located need to be updated, instead of the gains of all the remaining sensors, which will significantly reduce the calculation of the gains.
	This complete the proof.
\end{pf}

\subsection{The impact of different gain update methods on algorithm performance}
\textcolor{blue}{In this section, $5$ different gain update methods are compared in terms of computational complexity, distribution ratio vector accuracy, and performance. The considered methods are as follows:}
\begin{itemize}
	
	\item[(1)]\textcolor{blue}{Method $1$: Divide the total sensor set into $m_i$ subsets by Algorithm \ref{alg:3-4}, and update the gains of all fully correlated elements in the subset after each sensor selection.}
	
	\item[(2)]\textcolor{blue}{Method $2$: Divide the total sensor set into $m_i$ subsets by Algorithm \ref{alg:3-4}, and update the gains of all related elements after each sensor selection.}
	
	\item[(3)]\textcolor{blue}{Method $3$: Randomly divide the total sensor set into $m_i$ subsets, and  update the gains of all fully correlated sensors in the subset after each sensor selection.}
	
	\item[(4)]\textcolor{blue}{Method $4$: Randomly divide the total sensor set into $m_i$ subsets, and  update the gains of all related sensors in the subset after each sensor selection.}
	
	\item[(5)]\textcolor{blue}{Method $5$: without divide the totaol sensor set, and  update the gains of all related sensors after each sensor selection.}
\end{itemize}

%\textcolor{blue}{To ensure the fairness of the comparison, the following pre-set conditions are set: all methods that require set divide are divided into $m_i$ subsets, all estimated error vectors are processed by the de-influencing operation, the approach for selecting malicious information is based on the distribution ratio vector, and the attack intensity of each subset is completely balanced. The comparison results are shown in Table \ref{chapter3:tab compare}.}

\textcolor{blue}{To ensure fairness in the comparison, the following conditions are imposed. For all methods requiring partitioning, the total sensor set is divided into same number of subsets; all estimated error vectors are processed using the observable discrimination matrix; malicious information is selected based on the distribution ratio vector; and the attack intensity across subsets is  completely balanced. The comparison results are summarized in Table \ref{chapter3:tab compare}.}

\begin{table*}[!htb]
	\small
	\renewcommand{\arraystretch}{1.2}
	\centering
	\caption{\textcolor{blue}{Effect of different gain update methods on computational complexity, distribution ratio vector accuracy, and performance}}
	\label{chapter3:tab compare}
	\begin{tabular*}{1\textwidth}{@{\extracolsep{\fill}}cccccc}
		\hline
		% 第一行表头（把最后两列合并）
		%		& \begin{tabular}[c]{@{}c@{}}number of\\ subsets\end{tabular}
		%		& \begin{tabular}[c]{@{}c@{}}computational\\ complexity\end{tabular}
		%		& \begin{tabular}[c]{@{}c@{}}distribution ratio\\ vector accuracy\end{tabular}
		%		& \multicolumn{2}{c}{lower bound performance} \\
		
		&number of&computational&distribution ratio&\multicolumn{2}{c}{lower bound performance}\\
		\cline{5-6}
		% 第二行表头（拆分为 overall / subset）
		
		&
		subsets& complexity
		&  vector accuracy
		
		& overall 
		& subset \\
		\hline
		Method 1 & $m_i$ & larger                                                  & accurate                                                   & near-optimal     & $1-1/e$ \\
		Method 2 & $m_i$ & \begin{tabular}[c]{@{}c@{}}large\end{tabular}  & most accurate                                              & optimal  & $1-1/e$ \\
		Method 3 & $m_i$ & larger                                                  & random                                                     & random   & $1-1/e$ \\
		Method 4 & $m_i$ & \begin{tabular}[c]{@{}c@{}} larger\end{tabular} & \begin{tabular}[c]{@{}c@{}}random\end{tabular} & random   & $1-1/e$ \\
		Method 5 & 1     & largest                                                         & most accurate                                              & optimal, $1-1/e$ & none    \\
		\hline
	\end{tabular*}
\end{table*}
%\textcolor{blue}{Actually, the method $5$ is essentially the ADS algorithm proposed in the literature \cite{Suo2024Security}. Its computational complexity is the largest among the $5$ gain update methods, as this method requires normalization of high-dimensional vectors and updating the gains of all correlated elements. Therefore, the distribution ratio vector of this method is the most accurate and the overall performance is the best. In the following discussion, method $5$ will be regarded as the benchmark.}

\textcolor{blue}{Method $5$ essentially corresponds to the ADS algorithm proposed in \cite{Suo2024Security}. Among the five gain update methods, it exhibits the highest computational complexity, since it involves normalizing high-dimensional vectors and updating the gains of all correlated elements. Consequently, this method achieves the most accurate distribution ratio vector and delivers the best overall performance. In the subsequent discussion, Method $5$ will be used as the benchmark.}

%\textcolor{blue}{Theoretically, the distribution ratio vector accuracy and overall performance of methods $2$ and $5$ are close, but method $2$ has a smaller computational complexity, for example, the vector dimension of the normalization operation is lower. The main difference between methods $1$ and $2$ is whether to update the gains of the correlated elements in other subsets after each sensor selection. Therefore, compared with method $2$, method $1$ has a lower computational complexity. Correspondingly, the distribution ratio vector accuracy and performance of method $1$ are also reduced. The difference between methods $3$ and $4$ is mainly the difference in the number of updated element gains. Therefore, method $4$ has a larger computational complexity, but its distribution ratio vector accuracy is also higher.
%In addition, the subset performance of methods $1-4$ is guaranteed by theorem \ref{thm42}, that is, the theoretical lower bound is $1-1/e$. However, the overall performance of methods $1$ and $3-4$ is lower than that of method $5$, which is due to the inherent performance gap between centralized algorithms and distributed algorithms. It should be noted that, theoretically, method $1$ has better and more stable overall performance than random set dividing.}

\textcolor{blue}{Theoretically, methods $2$ and $5$ achieve comparable distribution ratio vector accuracy and overall performance, but method $2$ requires lower computational complexity (e.g., lower-dimensional normalization). The key distinction between methods $1$ and $2$ lies in whether correlated elements in other subsets are updated after each sensor selection; as a result, method $1$ has lower complexity but also reduced accuracy and performance. Similarly, the difference between methods $3$ and $4$ is the number of updated element gains, making method $4$ more complex but more accurate.  
	For methods $1$–$4$, subset-level performance is guaranteed by Theorem \ref{thm42}, with a theoretical lower bound of $1-1/e$. However, their overall performance remains inferior to method $5$, reflecting the inherent gap between distributed and centralized algorithms. Notably, method $1$ is expected to deliver better and more stable performance than random partitioning.}

\begin{rem}
\textcolor{blue}{	In fact, building on the proposed set partitioning strategy, the distributed structure is naturally amenable to parallel computing frameworks, as processing within each subset is independent and cross-subset synchronization is minimal, so further efficiency gains can be expected in practical implementations. Meanwhile, even without parallel acceleration, the per-iteration computational cost scales as $O(1/m)$ with the number of subsets $m$, where $m$ denotes the number of subsets.}
	
	\end{rem}

% \begin{multicols}{2}
	% \noindent Leibler divergence $D_{\mathrm{KL}}(P^*||P)>0$, therefore, $D_{\mathrm{KL}}(P^*||P)$ is a positive definite function.
	% Consider the Kullback-Leibler divergence as Lyapunov function $V$, then we have
	
	% \end{multicols}

% \leftsep
% \noindent
% \begin{equation*}
	% \begin{aligned}
		% \frac{\mathrm{d}V}{\mathrm{d}t}=&
		% \mathrm{tr}\left\{\left[\frac{\partial D_{\mathrm{KL}}(P^*||P)}{\partial \mathbf{p}_1},\frac{\partial D_{\mathrm{KL}}(P^*||P)}{\partial \mathbf{p}_2},\dots,\frac{\partial D_{\mathrm{KL}}(P^*||P)}{\partial \mathbf{p}_n}\right]
		% \left[\frac{\mathrm{d}\mathbf{p}_1}{\mathrm{d}t},\frac{\mathrm{d}\mathbf{p}_2}{\mathrm{d}t},\dots,\frac{\mathrm{d}\mathbf{p}_n}{\mathrm{d}t}\right]^\mathrm{T}\right\}\\
		% =&-\alpha\sum_{i=1}^n\left[\frac{\partial D_{\mathrm{KL}}(P^*||P)}{\partial \mathbf{p}_i}\right]^\mathrm{T}\left[\frac{\partial D_{\mathrm{KL}}(P^*||P)}{\partial \mathbf{p}_i}\right]<0\\
		% \end{aligned}\\
	% \end{equation*}
% \noindent
% \rightsep
% \begin{multicols}{2}
	% Therefore, as long as the $l-1=1$-th selected sensor $j_s^{(1)}$'s impact on the distribution ratio vector error of the $l=2$-th selected sensor $j_s^{(2)}$ is bounded, then the distribution ratio vector error is tolerable.
	% \end{multicols}

\section{Simulation Results}
In this section, simulation experiments are conducted on Automated Guided Vehicle (AGV) operation monitoring within an IoT scenario to illustrate the effectiveness of the proposed algorithm \citep{vlachos2024smart}.
In modern smart factories and warehouses, a large number of distributed sensors are deployed to continuously monitor the global operating states of AGVs, including their positions and velocities, in order to support real-time traffic management and scheduling.
The AGV dynamics can be described by the following discrete-time motion model \citep{zhou2022security},
	\begin{equation} \label{vehicle}
		\left[ \begin{array}{c}
			p_x\left( k+1 \right)\\
			p_y\left( k+1 \right)\\
			v_x\left( k+1 \right)\\
			v_y\left( k+1 \right)\\
		\end{array} \right] = \left[ \begin{matrix}
			1 & 0 & 1/50 & 0\\
			0 & 1 & 0 & 1/50\\
			0 & 0 & 1 & 0\\
			0 & 0 & 0 & 1\\
		\end{matrix} \right] \left[ \begin{array}{c}
			p_x\left( k \right)\\
			p_y\left( k \right)\\
			v_x\left( k \right)\\
			v_y\left( k \right)\\
		\end{array} \right] + \omega \left( k \right),
	\end{equation}
where $p_x\left( k \right)$, $p_y\left( k \right)$, $v_x\left( k \right)$, and $v_y\left( k \right)$ represent the position and velocity of the vehicle in the $x$ and $y$ directions at moment $k$, respectively.  
The initial state of the vehicle is $x(0) = \left[ \begin{matrix} 50 & 0 & 5 & 0 \end{matrix} \right]^T$.  
% In an ideal scenario, the vehicle would move in a uniform straight line along the $x$-axis direction. 
In the subsequent simulations, the initial state estimates of all sensors, $\hat{x}_{i}(0)$, are set to $x(0)$.

\textcolor{blue}{This simulation considers 2 kinds of sensor network scenarios to validate the performance of the proposed algorithm}
\begin{enumerate}
	\item[1)] \textcolor{blue}{Scenario 1: A single central sensor and 100 neighboring sensors in an ultra-large-scale network.}
	\item[2)] \textcolor{blue}{Scenario 2: A complex distributed network with 500 sensors.}
\end{enumerate}

\textcolor{blue}{And all simulations were conducted on a personal laptop equipped with an Intel i7-13700H CPU, 32 GB of RAM, running MATLAB R2022b.}

\textbf{Scenario 1}: 
In this scenario, a central sensor (labeled 0) is considered, along with 100 neighboring sensors (labeled 1–100). Each sensor independently measures the state of the vehicle, with the measurement model given by:
\begin{equation}
	y_{i}\left( k \right) =C_{i}x\left( k \right) +\nu _{i}\left( k \right).
\end{equation}
where the measurement matrix for the central sensor is $C_0 = [ \begin{matrix} 1 & 0 & 0 & 0 \end{matrix} ]$.  
The measurement matrices for the neighboring sensors differ based on their observable state space, and are given by the following four types:
	\begin{eqnarray} \label{obse_matrix}
		C_1&=&[ \begin{matrix}
			1&		0&		0&		0\\
		\end{matrix} ]
		,\quad
		C_2=[ \begin{matrix}
			0&		1&		0&		0\\
		\end{matrix} ]
		, \nonumber\\
		C_3&=&[ \begin{matrix}
			0&		0&		1&		0\\
		\end{matrix} ]
		,\quad
		C_4=[ \begin{matrix}
			0&		0&		0&		1\\
		\end{matrix} ].
		%  \nonumber\\
		% C_{5}&=&\left[ \begin{matrix}
			% 		0&		0&		1&		0\\
			% 		0&		0&		0&		1\\
			% 	\end{matrix} \right]
		%  ,\quad
		%  C_{6}=\left[ \begin{matrix}
			% 		1&		0&		0&		0\\
			% 		0&		1&		0&		0\\
			% 	\end{matrix} \right]
	\end{eqnarray}
The process noise $\omega$ and measurement noise $\nu_i$ are set with parameters $Q = 0.5I$ and $R_i = 0.5I$, respectively. Additionally, both the process noise and the measurement noise are bounded, i.e., $||\omega(k)||_{\infty} \leq 0.05$ and $||\nu_i(k)||_{\infty} \leq 0.05$ at all times. The observability matrices for each sensor are given by:
	\begin{eqnarray}
		&&Q_{1,o} = \left[ \begin{matrix}  
			1 & 1 & 1 & 1 \\  
			0 & 0 & 0 & 0 \\  
			0 & 1 & 2 & 3 \\  
			0 & 0 & 0 & 0  
		\end{matrix} \right], \quad  
		Q_{2,o} = \left[ \begin{matrix}  
			0 & 0 & 0 & 0 \\  
			1 & 1 & 1 & 1 \\  
			0 & 0 & 0 & 0 \\  
			0 & 1 & 2 & 3  
		\end{matrix} \right],\nonumber\\
		&&Q_{3,o} = \left[ \begin{matrix}  
			0 & 0 & 0 & 0 \\  
			0 & 0 & 0 & 0 \\  
			1 & 1 & 1 & 1 \\  
			0 & 0 & 0 & 0  
		\end{matrix} \right], \quad  
		Q_{4,o} = \left[ \begin{matrix}  
			0 & 0 & 0 & 0 \\  
			0 & 0 & 0 & 0 \\  
			0 & 0 & 0 & 0 \\  
			1 & 1 & 1 & 1  
		\end{matrix} \right].
	\end{eqnarray}

It can be seen that the observability dimensions for the different types of sensors are varied, with dimensions of 1 and 3, 2 and 4, 3, and 4, respectively. It is assumed that the measurement matrix of sensors 1–25 is  $C_1$, the measurement matrix of sensors 26–50 is $C_2$, the measurement matrix of sensors 51–75 is $C_3$, and the measurement matrix of sensors 76–100 is $C_4$. Additionally, all neighboring sensors $j \in \mathcal{N}_0$ are capable of transmitting information to the central sensor, with the transmitted information being the state estimate $\hat{x}_{0,j}(k)$, which may be compromised by FDIAs. Based on the set partitioning strategy in Algorithm \ref{alg:3-4}, the 100 sensors are divided into four subsets: sensors 1–25 form subset 1, sensors 26–50 form subset 2, sensors 51–75 form subset 3, and sensors 76–100 form subset 4.

\textcolor{blue}{At any moment, the attacker randomly selects 40 sensors to attack from  100 neighboring sensors. For any consecutive moments, the sets of attacked sensors $\mathcal{A}_{0,\,k}$ and $\mathcal{A}_{0,\,k-1}$ in the neighborhood of sensor 0 satisfy $\varDelta _{0,\,k}=( \mathcal{A}_{0,\,k}\backslash \mathcal{A}_{0,\,k-1} ) \cup ( \mathcal{A}_{0,\,k-1}\backslash \mathcal{A}_{0,\,k} ) \ge 0$.}

In this simulation, only the velocity estimate $\hat{x}_{0,\,j}(k)$ in the sensor's estimated results is tampered with by the attacker, while the position estimate remains unaffected. \textcolor{blue}{The considered attack signals includes two types: one type corresponds to a unstealthy attack, which exhibits large magnitudes most of the time; the other type corresponds to a stealthy attack, whose magnitude is close to the estimation noise most of the time.} To better simulate different attack scenarios, the attack signals are designed based on the magnitude of the estimation noise, ensuring that stealthy and unstealthy attacks do not appear in the same simulation experiment. This setup meets the prerequisites of Lemma 2 shown in the
APPENDIX A.

According to Kalman decomposition \citep{Suo2024attack}, by first deinfluence and then reconstruction, the estimated error $\Delta\hat{x}_j^{re}$ of each sensor $i$ and its neighbor sensor $j\in\mathcal{N}_i$, which only contains the observable part, can be obtained. The augmented error matrix $\varLambda_{i,k}$ in equation (\ref{f_A}) is obtained by summarizing the $\Delta\hat{x}_j^{re}$ of all neighbors $j$. It should be noted that the number of non-zero rows of $\Delta\hat{x}_j^{re}$ and $Q_{j,\,o}$ is the same.

% \hl{Add to affect the results of the operation}

% \begin{definition}(\textbf{Estimation Error Reconstruction})
	%     At time $k$, the error vector $\Delta\hat{x}_{j}(k)$ between the estimated value $\hat{x}_i(k)$ of sensor $i$ and the estimated value $\hat{x}_{ij}(k)$ of neighbor sensor $j$, $j\in\mathcal{N}_i$, where the part that can be observed by the observation matrix $C_j$ is modeled as:
	% \begin{equation} \label{error_between_estimate_and_actual}
		% \varDelta \hat{x}_{j,o}\left( k \right) =\hat{x}_{ j,o }\left( k \right) -\hat{x}_{ i,o }\left( k \right) =V_{j,o}^{\mathrm{T}}\left( \hat{x}_{ij}\left( k \right) -\hat{x}_i\left( k \right) \right).
		% \end{equation}
	% The error after reconstruction is $\varDelta \hat{x}_{j}^{re}\left( k \right)=V_{j,o}\varDelta \hat{x}_{j,o}\left( k \right)$, where $V_{j,o}$ comes from the linear non-singular transformation $F_i=[ V_{i,\Bar{o}}\,\,V_{i,o} ] ^{\mathrm{T}}$, which transforms the system state equation (\ref{chapter2:state_equation}) into $\Bar{x}_i\left( k+1 \right) =F_iA F_{i}^{\mathrm{T}}\Bar{x}_i\left( k \right)+F_i\omega \left( k \right)$.
	% \end{definition}

Next, we compare the distribution of inter-subset gain mutual influence and intra-subset correlation under different partitioning strategies. Three partitioning strategies are considered: the set partitioning strategy based on minimizing inter-subset gain mutual influence (Partitioning Strategy $1$), the  set partitioning strategy based on the Grassmann distance (Partitioning Strategy $2$), and the random division of $100$ sensors into $4$ groups (Partitioning Strategy $3$).  
For Partitioning Strategy $1$, the observable spaces of $C_1$ ($C_2$) and $C_3$ ($C_4$) overlap. Consequently, the partitioning result is that subset $1$ includes all sensors with measurement matrices $C_1$ and $C_3$, while subset $2$ includes all sensors with measurement matrices $C_2$ and $C_4$. As shown in TABLE \ref{chapter3:tab:min_wai_max_nei}, the inter-subset benefit influence is $0$, while the intra-subset correlation is $0.5$, indicating that this strategy minimizes the inter-subset gain mutual influence.  
For Partitioning Strategy $2$, sensors are fully correlated only when their measurement matrices are identical. Thus, the partitioning result consists of 4 subsets, each containing one type of sensor. As shown in TABLE \ref{chapter3:tab:min_wai_max_nei}, the inter-subset gain mutual influence is $625$, while the intra-subset correlation is $1$, indicating that this strategy maximizes intra-subset correlation.  
For Partitioning Strategy $3$, by averaging the results of $100$ completely independent partitioning instances, we obtain the results shown in TABLE \ref{chapter3:tab:min_wai_max_nei}, where neither the inter-subset gain mutual influence nor the intra-subset correlation is optimal.  
Therefore, the proposed set partitioning strategy achieves the goal of minimizing inter-subset gain mutual influence while maximizing intra-subset correlation.

\begin{table*}[ht]
	%\small
	\renewcommand{\arraystretch}{1.3}
	\centering
	\caption{The inter-subset interaction and intra-subset correlation under different partitioning strategy}
	\label{chapter3:tab:min_wai_max_nei}
	\begin{tabular*}{0.9\textwidth}{@{\extracolsep{\fill}}cccccccc}
		
		\toprule
		Strategy & Set  & Subset 1                         & Subset 2                         & Subset 3                         & Subset 4  & \multicolumn{2}{c}{ Intra-subset Correlation}                       \\ \midrule
		\multirow{2}{*}{Strategy 1} & Subset 1& \emptyDiag{11581}& 0 &\emptyDiag{11581} &\emptyDiag{11581} & Subset 1 & 0.5\\
		& Subset 2& 0&\emptyDiag{11581} &\emptyDiag{11581} &\emptyDiag{11581} & Subset 2 & 0.5
		\\\bottomrule
		%\midrule
		\multirow{4}{*}{Strategy 2} & Subset 1    &         \emptyDiag{11581}                    & 0   & 625 & 0 & Subset 1 & 1  \\ 
		&Subset 2    & 0   &               \emptyDiag{11581}                & 0   & 625 & Subset 2 & 1  \\ 
		&Subset 3    & 625 & 0   &                        \emptyDiag{11581}       & 0  & Subset 3 & 1 \\
		&Subset 4    & 0   & 625 & 0   &                \emptyDiag{11581}     & Subset 4 & 1          \\\bottomrule
		%     \end{tabular*}
	% \end{table*}

% \begin{table*}[ht]
	%     \small
	%     \renewcommand{\arraystretch}{1.4}
	%     \centering
	%     \caption{Subset Inter-benefit Interaction and Internal Correlation under Grouping Strategy 3}
	% \label{chapter3:tab:random}
	% \begin{tabular*}{0.9\textwidth}{@{\extracolsep{\fill}}ccccccc}
		%     \toprule
		% Subset Inter-benefit Interaction & Subset 1                         & Subset 2                         & Subset 3                         & Subset 4   & \multicolumn{2}{c}{Subset Internal Correlation}                        \\ \cline{1-5} \cline{6-7}
		\multirow{4}{*}{Strategy 3} &Subset 1                         & \emptyDiag{11581}                & 158  & 160  & 167  & Subset 1 & 0.293 \\ 
		&Subset 2                         & 158  & \emptyDiag{11581}             & 140  & 156  & Subset 2 & 0.28  \\ 
		&Subset 3                         & 160  & 140  & \emptyDiag{11581}           & 170  & Subset 3 & 0.306 \\
		& Subset 4                         & 167  & 156  & 170  & \emptyDiag{11581} & Subset 4 & 0.261 \\\bottomrule
	\end{tabular*}
\end{table*}

Again, to verify the conclusion of Theorem \ref{thm48}, this simulation considers a special scenario where, at the first selection of each moment, the D-ADS algorithm always selects an attacked sensor from the subset $4$. 
%Two types of gain update strategies are considered.
%, corresponding to gain update method $1$ and method $2$ as described in Section 4.3.4.  
% 1. Updating only the gains of all remaining elements in subset $4$, 2. Updating the gains of all remaining elements in subsets $2$ and $4$.  

\begin{figure}[htb]
	\centering
	\includegraphics[width=0.6\textwidth]{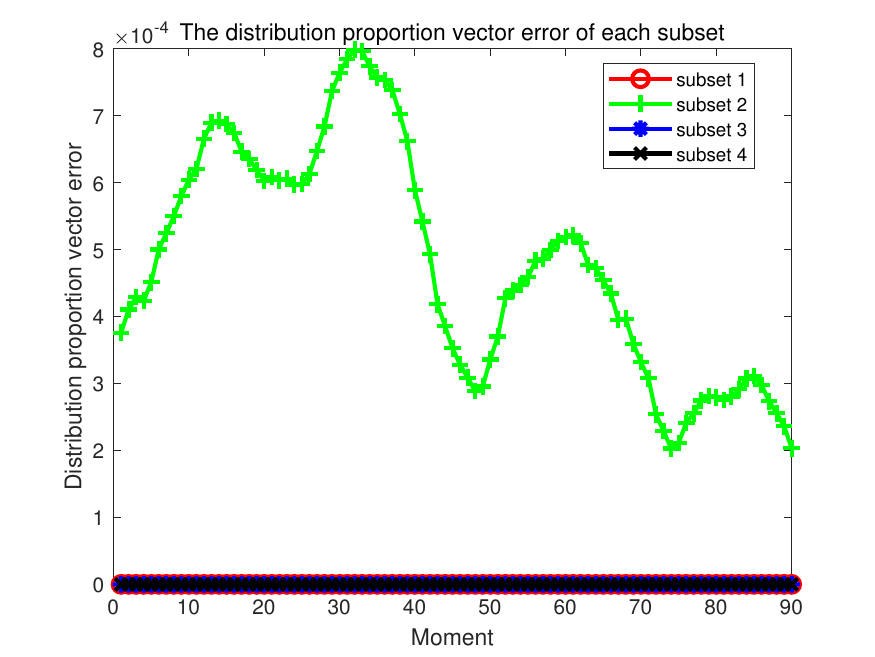}
	\caption{Distribution ratio vector error of the proposed partitioning strategy under different gain update methods}
	\label{chapter3:fig:p_error}
\end{figure}

According to Theorem \ref{thm48}, the gain interactions only exist  between $C_2$ and $C_4$, that is, under these two gain update strategies, the error in the distribution ratio vector only appears in subset $2$. The error curves are shown in Figure \ref{chapter3:fig:p_error}. To make the error curves smoother and more intuitive, the evaluation metric used is windowed RMSEs (W-RMSEs), with a window size of $10$ moments. It can be observed that the distribution ratio vector error in subset $2$ is bounded, with an average value of only $2.2871e-04$ over the entire time period, while the errors in subsets $1$, $3$, and $4$ remain zero throughout, which is consistent with the conclusion of Theorem \ref{thm48}.

\textcolor{blue}{To more comprehensively evaluate the performance of the proposed algorithm,  the absolute differences between D-ADS and the baseline ADS under different sensor selection methods is compared. The evaluation metric used is the average optimization rate defined in Definition \ref{chapter2:optimization_rate_index}, which is extended to the distributed case.}
	
\begin{defn}\textbf{(Average Optimization Rate)} \citep{Suo2024Security}\label{chapter2:optimization_rate_index}
	\textcolor{blue}{For the entire time period $T$, the average optimization rate of the $i$-th sensor is defined as the ratio of the \emph{objective function} (\ref{f_A}) value of the candidate set $\mathcal{A} _{i,k} $ (or $\mathcal{A} _{i,\,k,\,g_i} $) selected by Algorithm \ref{alg:3-1}, which is defined as $\frac{1}{T}\sum_{k=1}^T{( f_k( \mathcal{A} _{i,k} ) /f_k( \mathcal{A} _{i,k}^{*} ) )}$ (or $\frac{1}{T}\sum_{k=1}^T{( f_k( \mathcal{A} _{i,\,k,\,g_i} ) /f_k( \mathcal{A} _{i,\,k,\,g_i}^{*} ) )}$). And the average optimization rate for all the subset $g_i=1:m_i$ can be defined as  $\frac{1}{T\cdot m_i}\sum_{k=1}^T\sum_{g_i=1}^{m_i}{( f_k( \mathcal{A} _{i,\,k,\,g_i} ) /f_k( \mathcal{A} _{i,\,k,\,g_i}^{*} ) )}$.}
\end{defn}
%		\begin{defn}\textbf{(Average Optimization Rate)} \citep{Suo2024Security}\label{chapter2:optimization_rate_index}
%			\textcolor{blue}{At moment $k$, the average optimization rate of the $i$-th sensor is defined as the ratio of the \emph{objective function} (\ref{f_A}) value of the candidate set $\mathcal{A} _{i,k} $ (or $\mathcal{A} _{i,\,k,\,g_i} $) selected by Algorithm \ref{alg:3-1}, which is defined as $f_k( \mathcal{A} _{i,k}) /f_k( \mathcal{A} _{i,k}^{*} )$ (or $f_k( \mathcal{A} _{i,\,k,\,g_i}) /f_k( \mathcal{A} _{i,\,k,\,g_i}^{*} )$) . And for the entire time period $T$, the average optimization rate can be defined as $\frac{1}{T}\sum_{k=1}^T{( f_k( \mathcal{A} _{i,k} ) /f_k( \mathcal{A} _{i,k}^{*} ) )}$ (or $\frac{1}{T\cdot m_i}\sum_{k=1}^T\sum_{g_i=1}^{m_i}{( f_k( \mathcal{A} _{i,\,k,\,g_i} ) /f_k( \mathcal{A} _{i,\,k,\,g_i}^{*} ) )}$).}
%		\end{defn}

	 \textcolor{blue}{Based on the set partitioning strategy proposed in this paper for partitioning $100$ sensors and applying gain update method $1$ described in Section 3.4 to update element gains, we compare the average optimization rate for different sensor selection methods across each subset and the overall set. Theorem \ref{thm42} states that the theoretical lower bound on the average optimization rate for any subset under the proposed D-ADS algorithm is $1 - 1/e$. }

\textcolor{blue}{The two sensor selection methods considered are the probabilistic selection method and the ranking selection method. The probabilistic selection method employs a probabilistic approach in the first stage, followed by a uniform random selection algorithm in the second stage, whereas the ranking selection method adopts a ranking-based approach in the first stage, followed by a uniform random selection algorithm in the second stage. As shown in Table \ref{tab4-6}, the average optimization rate for individual subsets is unstable, typically fluctuating around the overall set’s average optimization rate. This is because it is difficult to ensure that the attack intensity is perfectly balanced across all subsets. Moreover, regardless of whether it is a subset or the overall set, the ranking selection method always achieves a higher average optimization rate than the probabilistic selection method. As shown in Table \ref{tab_performance_compare}, the absolute difference in average optimization rate between ADS and D-ADS shows that the performance of the ADS algorithm is almost always better than that of the D-ADS algorithm in different situations. However, the difference is very limited, with the maximum absolute difference being $1.648\%$, which is tolerable. }

\textcolor{blue}{Additionally, comparing the malicious information selection performance of the proposed algorithm under unstealthy and stealthy attacks, it is evident that the average optimization rate under unstealthy attacks is always higher than that under stealthy attacks.}

\textcolor{blue}{Finally, to demonstrate the generality of the proposed algorithm, we further investigate the combined impact of different subset numbers and attacked sensor dimensions (i.e., the number of attacked sensors) under the case of 100 sensors, and analyze their effects on both the average number of attacked subsets and the fraction of attacked subsets. Specifically, the number of subsets is set to $[2, 4, 5, 10, 20, 30, 40, 50]$, and the attacked sensor dimensions vary from 1 to 100 with a step size of 5.}

\begin{figure}[htb] \centering \begin{minipage}{\linewidth} \subfigure[Number of attacked subsets]{ \includegraphics[width=0.46\linewidth]{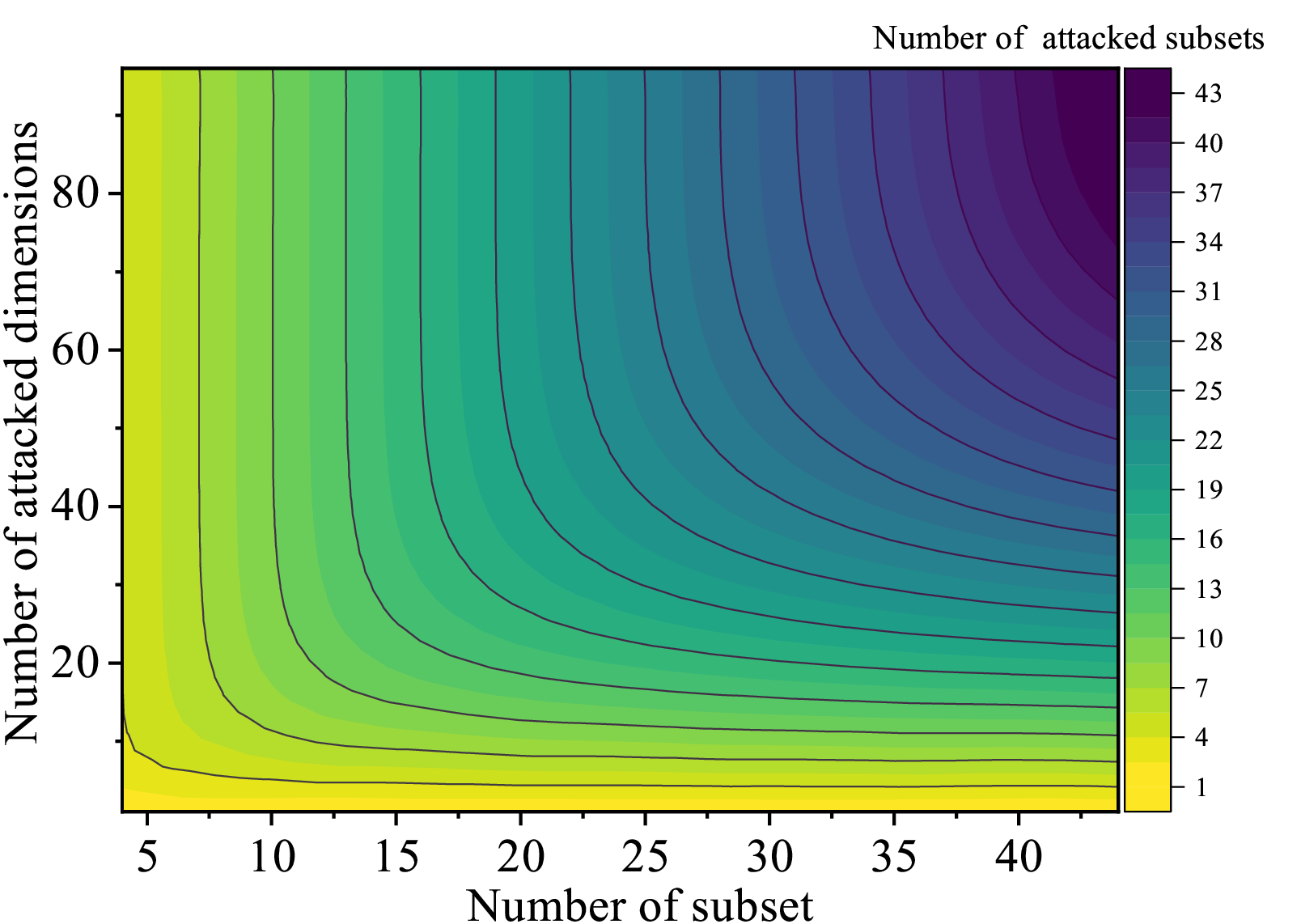} } \subfigure[Fraction of attacked subsets]{ \includegraphics[width=0.46\linewidth]{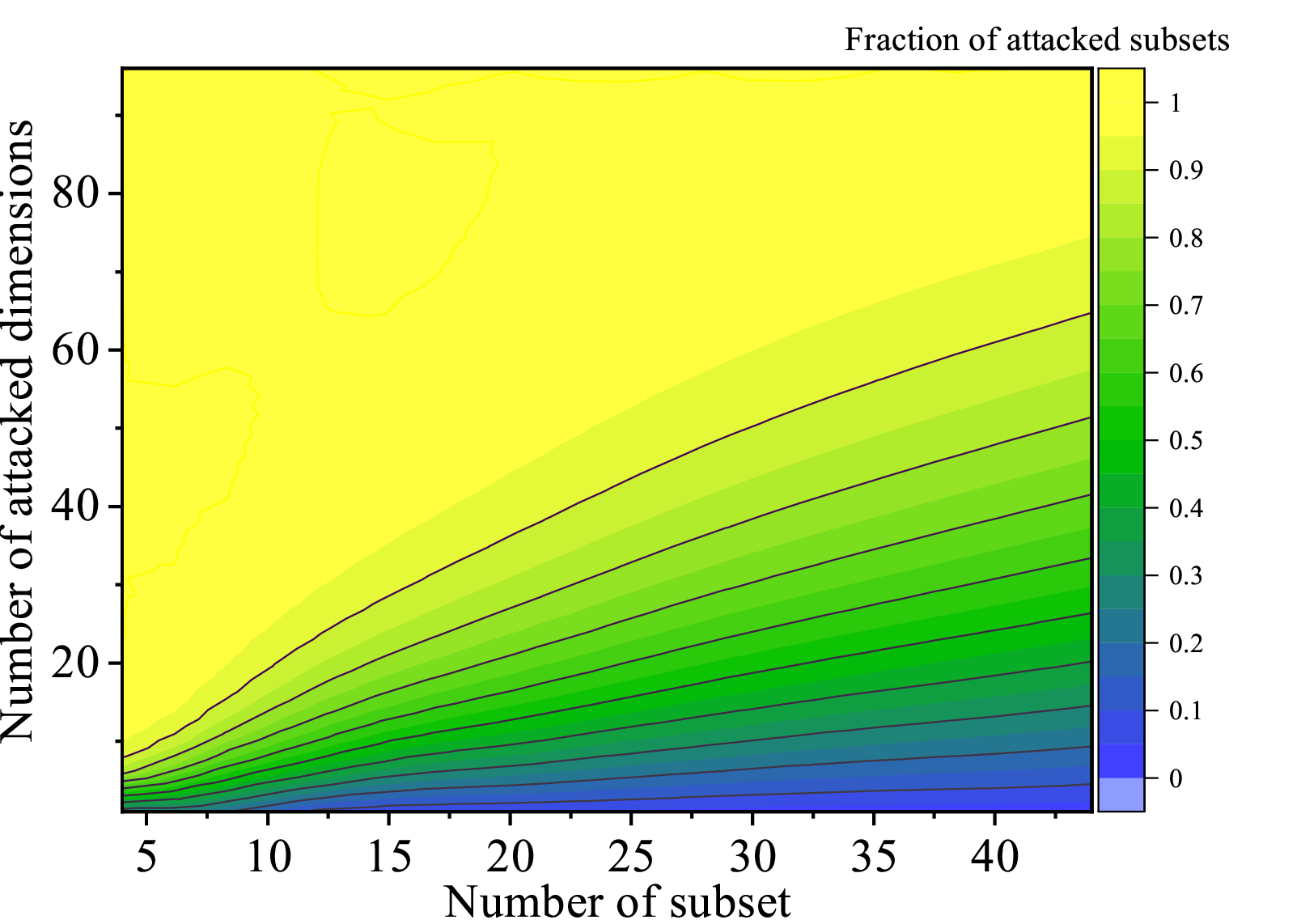} } \caption{\textcolor{blue}{The combined impact of different subset numbers and attacked sensor dimensions}} \label{ES_and_Frac} \end{minipage} \end{figure}

\textcolor{blue}{The simulation results are shown in Figure \ref{ES_and_Frac}. As illustrated in Figure \ref{ES_and_Frac}(a), the number of attacked subsets increases with both the number of subsets and the attacked dimensions. In Figure \ref{ES_and_Frac}(b), the fraction of attacked subsets also increases as the attacked dimensions grow. In the previous simulation, 100 sensors were divided into 4 subsets with 40 attacked dimensions, leading to all subsets being affected. However, the results in Figure \ref{ES_and_Frac} suggest that dividing the sensors into more subsets can significantly reduce the fraction of attacked subsets. This implies that, by ensuring independence or weak correlations among subsets, the number of subsets requiring gain updates can be reduced, thereby lowering the overall computational cost.}

\textcolor{blue}{For example, when the sensors are divided into 20 subsets, each containing 5 sensors, and all subsets are mutually independent or weakly correlated, with 40 attacked dimensions, on average only about 13 subsets are affected. According to the proposed gain update Method 1, it is then sufficient to update only those 13 subsets (85 sensors in total), while the remaining subsets remain untouched. Compared with the previous case of 4 subsets, this reduces the computational cost by approximately $15\%$.}

%\begin{table}[ht]
%	%\small
%	\renewcommand{\arraystretch}{1.2}
%	\caption{\textcolor{blue}{Average Optimization Rate of Each Subset and the Total Set under Different Sensor Selection Methods}}
%	\label{tab4-6}
%	\centering
%	\begin{tabular*}{0.9\textwidth}{@{\extracolsep{\fill}}ccccc}
%		\toprule
%		\multirow{2}{*}{Set} & \multicolumn{2}{c}{ Selection Method} & \multicolumn{2}{c}{Average Optimization Rate} \\
%		& Probability & Ranking & Unstealthy & Stealthy \\ \midrule
%		\multirow{2}{*}{Subset 1} & $\checkmark$ & & 0.654 & 0.482 \\
%		& & $\checkmark$ & 0.817 & 0.830 \\
%		\multirow{2}{*}{Subset 2} & $\checkmark$ & & 0.763 & 0.672 \\
%		& & $\checkmark$ & \textbf{0.932} & 0.782 \\
%		\multirow{2}{*}{Subset 3} & $\checkmark$ & & 0.713 & \textbf{0.717} \\
%		& & $\checkmark$ & 0.882 & \textbf{0.899} \\
%		\multirow{2}{*}{Subset 4} & $\checkmark$ & & \textbf{0.821} & 0.658 \\
%		& & $\checkmark$ & 0.920 & 0.703 \\
%		\multirow{2}{*}{Ave Subset(D-ADS)} & $\checkmark$ & & 0.738 & \textbf{0.632} \\
%		& & $\checkmark$ & 0.888 & 0.804 \\
%		\multirow{2}{*}{Total(ADS)} & $\checkmark$ & & 0.745 & 0.622 \\
%		& & $\checkmark$ & 0.893 & 0.812 \\ \bottomrule
%	\end{tabular*}
%\end{table}

\begin{table}[ht]
	%\small
	\renewcommand{\arraystretch}{1.2}
	\caption{\textcolor{blue}{Average optimization rate of each subset and the total set with different sensor selection methods}}
	\label{tab4-6}
	\centering
	\begin{tabular*}{1\textwidth}{@{\extracolsep{\fill}}cccccc}
		\toprule
		\multirow{2}{*}{Algorithm}&	\multirow{2}{*}{Set} & \multicolumn{2}{c}{ Selection Method} & \multicolumn{2}{c}{Average Optimization Rate} \\\cline{3-4}\cline{5-6}
		&	& Probability & Ranking & Unstealthy & Stealthy \\ \midrule
		\multirow{10}{*}{D-ADS}&	\multirow{2}{*}{Subset 1} & $\checkmark$ & & 0.654 & 0.482 \\
		&	& & $\checkmark$ & 0.817 & 0.830 \\ \cline{2-6}
		&	\multirow{2}{*}{Subset 2} & $\checkmark$ & & 0.763 & 0.672 \\
		&	& & $\checkmark$ & {0.932} & 0.782 \\\cline{2-6}
		&	\multirow{2}{*}{Subset 3} & $\checkmark$ & & 0.713 & {0.717} \\
		&	& & $\checkmark$ & 0.882 & {0.899} \\\cline{2-6}
		&	\multirow{2}{*}{Subset 4} & $\checkmark$ & & {0.821} 	& 0.658 \\
		&	& & $\checkmark$ & 0.920 & 0.703 \\\cline{2-6}
		&	\multirow{2}{*}{Average Subset} & $\checkmark$ & & 0.738 & {0.632} \\
		&	& & $\checkmark$ & 0.888 & 0.804 \\ \midrule
		\multirow{2}{*}{ADS}&	\multirow{2}{*}{Total Set} & $\checkmark$ & & 0.745 & 0.622 \\
		&	& & $\checkmark$ & 0.893 & 0.812 \\ \bottomrule
	\end{tabular*}
\end{table}

\begin{table}[ht]
	\renewcommand{\arraystretch}{1.2}
	\caption{\textcolor{blue}{The absolute difference in average optimization rate between ADS and D-ADS algorithms in different situations}}
	\label{tab_performance_compare}
	\centering
	\begin{tabular*}{0.5\textwidth}{@{\extracolsep{\fill}}ccc}
		\toprule
		& Unstealthy & Stealthy \\\midrule
		Probability & 0.973$\%$    & 1.648$\%$   \\
		Ranking     & 0.588$\%$     & 1.047$\%$  \\
		\bottomrule
	\end{tabular*}
\end{table}

%\begin{table}[ht]
%	\renewcommand{\arraystretch}{1.2}
%	\caption{\textcolor{blue}{The difference in average optimization rate between ADS and D-ADS algorithms in different situations}}
%	\label{tab_performance_compare}
%	\centering
%	\begin{tabular*}{0.5\textwidth}{@{\extracolsep{\fill}}ccc}
%		\toprule
%		& Unstealthy & Stealthy \\\midrule
%		Probability & -0.973$\%$    & 1.648$\%$   \\
%		Ranking     & -0.588$\%$     & -1.047$\%$  \\
%		\bottomrule
%	\end{tabular*}
%\end{table}

% \begin{table*}[ht]
	% 	\small
	% 	\renewcommand{\arraystretch}{1.2}
	% 	\caption{Average Optimization Rate of Each Subset and the Total Set under Different Sensor Selection Methods}
	% 	\label{tab4-6}
	% 	\centering
	% 	\begin{tabular*}{0.8\textwidth}{@{\extracolsep{\fill}}ccccc}
		% 		\toprule
		% 		\multirow{2}{*}{Set} & \multicolumn{2}{c}{Sensor Selection Method} & \multicolumn{2}{c}{Average Optimization Rate} \\
		% 		& Probability-based Selection & Ranking-based Selection & Non-hiding Attack & Hiding Attack \\ \midrule
		% 		\multirow{2}{*}{Subset 1} & $\checkmark$ & & 0.654 & 0.482 \\
		% 		& & $\checkmark$ & 0.817 & 0.830 \\
		% 		\multirow{2}{*}{Subset 2} & $\checkmark$ & & 0.763 & 0.672 \\
		% 		& & $\checkmark$ & 0.932 & 0.782 \\
		% 		\multirow{2}{*}{Subset 3} & $\checkmark$ & & 0.713 & 0.717 \\
		% 		& & $\checkmark$ & 0.882 & 0.899 \\
		% 		\multirow{2}{*}{Subset 4} & $\checkmark$ & & 0.821 & 0.658 \\
		% 		& & $\checkmark$ & 0.920 & 0.703 \\
		% 		\multirow{2}{*}{Total Set} & $\checkmark$ & & 0.745 & 0.622 \\
		% 		& & $\checkmark$ & 0.893 & 0.812 \\ \bottomrule
		% 	\end{tabular*}
	% \end{table*}

\textbf{Scenario 2}: Next, this simulation verifies that the proposed algorithm can ensure the secure state estimation of the system.  
Consider a complex distributed network scenario in which $500$ sensors (labeled $1-500$) are randomly deployed in a $200$m $\times 200$m area. Each sensor has the capability to communicate with sensors within a $30$m radius, forming an undirected graph. Similar to Scenario $1$, each sensor can independently measure the state of the vehicles, and the measurement matrix is classified into the $4$ types given in equation (\ref{obse_matrix}).  
First, measurement matrices are randomly assigned to the $500$ sensors, and the number of each type of sensor in the neighbor set of each sensor is recorded. The statistical results indicate that, for almost all sensors, their neighbor sets fail to satisfy the requirement stated in Remark \ref{chapter3:remark4.1}, which mandates an equal number of sensors observing different state spaces.  
% In this case, the proposed Algorithm 3.4 as shown in the APPENDIX C of the full version of this paper \hl{[]} is applied to the entire network to dynamically adjust the communication directions between sensors, thereby controlling the distribution of neighbor sensors for each sensor. 
In this case, based on the set partitioning result, the communication direction between sensors is dynamically adjusted to control the distribution of neighbor sensors of each sensor. Ultimately, the entire network is transformed into a directed graph, ensuring that the neighbor set of each sensor satisfies Remark \ref{chapter3:remark4.1}.  
Therefore, based on the proposed Algorithm \ref{alg:3-3} or Algorithm \ref{alg:3-4}, the neighbor set of each sensor can be divided into $4$ subsets.
\begin{figure}[htb]
	\centering
	\includegraphics[width=0.6\textwidth]{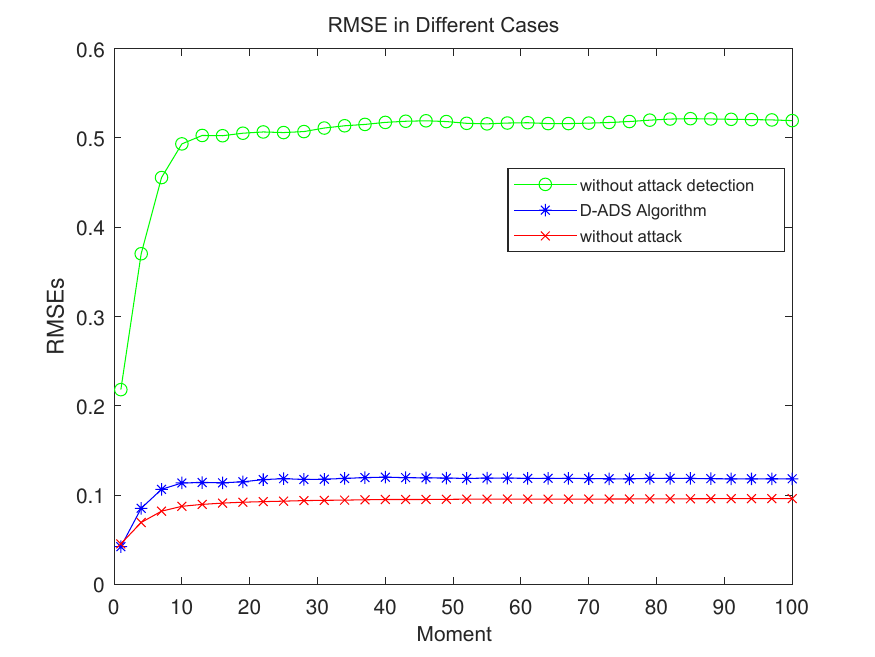}
	\caption{RMSE curves under different cases}
	\label{chapter3:fig:RMSE_distribued}
\end{figure}

\textcolor{blue}{At each moment $k=1:100$, the attacker randomly selects no more than half of the target sensors from the set of neighboring sensors of each sensor and launches stealthy attacks on their communication links.} Therefore, the estimation results transmitted between sensors may be affected by FDIAs. Based on the D-ADS Algorithm, each sensor can preliminarily exclude a malicious information set before fusing the information from neighboring sensors.
This simulation compares the security of the system under different cases, including the case without attack detection, the case using the proposed D-ADS algorithm, and the case without any attack. For the no-attack scenario, the mean of the estimation errors of all sensors at each moment is used to calculate the RMSE. In contrast, for the case where attacks exist, considering that the security of the system can be intuitively described by the maximum estimation error, this simulation calculates the RMSE using the maximum estimation error at each moment.
Figure \ref{chapter3:fig:RMSE_distribued} presents the RMSE curves under these cases, where there are three curves in total: the no-attack case (red), the proposed D-ADS algorithm detection (blue), and the case without  attack detection (green). It can be seen that the blue curve is significantly lower than the green curve, indicating that the proposed algorithm ensures system security. However, there is still a certain gap between the blue and red curves, as stealthy attacks are challenging to detect accurately. This observation is consistent with the simulation results in Scenario $1$.

\vspace{-3mm}
\section{Conclusions}
\textcolor{blue}{Facing the impact of the expanding scale of IoT sensor networks on the efficiency of malicious information selection tasks, this paper investigates the application potential of distributed algorithms in large-scale IoT from the perspective of set partitioning strategy. Theoretical analysis and simulation results demonstrate that the proposed set partitioning strategy effectively narrows the performance gap between distributed and centralized methods, while ensuring that computational cost decreases as the number of subsets increases. The findings indicate that the proposed algorithm serves as an efficient cost-reduction strategy for detection algorithms, enabling them to be more feasible for resource-constrained edge or gateway devices in the Iot, while preserving detection reliability, thereby providing an efficient and feasible security protection path for practical IoT systems.
	Future research will explore more advanced feature extraction methods to capture the complex and dynamic information in IoT networks, thereby addressing sophisticated attacks such as Advanced Persistent Threats (APTs) more effectively.}

%\textcolor{blue}{This paper investigates the performance of distributed attack detection scheduling algorithms in large-scale IoT networks. By analyzing the gain mutual influence between sensor subsets, an set partitioning strategy based on Grassmann distance is proposed, effectively reducing inter-subset mutual influence while enhancing intra-subset correlation. Theoretical analysis and simulation results demonstrate that the proposed strategy not only significantly narrows the performance gap between the D-ADS and ADS algorithm but also reduces computational cost. Future research will explore more advanced feature extraction methods to better capture complex and dynamic information within IoT networks. This direction is expected to further enhance the efficiency and accuracy of attack detection, especially under large-scale and adversarial scenarios. }

%Future research can explore more advanced feature extraction methods to better capture complex information within the network, thereby further improving the efficiency and accuracy of attack detection.

\vspace{10mm}
\appendix
%\appendices
%\appendices
\section{APPENDIX A: The balance of the attack strategy}

According to Assumption \ref{chapter2:attack_assum}, at each moment, the attacker randomly selects $q_i$ sensors from the neighbor set $\mathcal{N}_i$ of sensor $i$ to launch the FDIAs, and the attack strategy satisfies the dynamic attack strategy in Definition \ref{semi-dynamic attack defn}.  Lemma \ref{lem41} will analyze the balance between the number of attacked sensors and the attack intensity in each subset.
\begin{lem}\label{lem41}
    Based on the Assumption \ref{chapter2:attack_assum}, during the entire time period, the expectation of the number of attacked sensors in the $g_i$-th subset $\mathcal{N} _{i,\,g_i}$ of sensor $i$ is $\mathbb{E}[ q _{i,\,g_i}]= \lfloor q_i/m_i \rfloor$. In addition, the intensity of the attack on each subset is also balanced. Both are independent of the sensor set partitioning strategy, provided that the attacker selects the sensor to be attacked completely randomly at each moment, and the cardinality of each subset is approximately average.
\end{lem}
% \begin{proof}
%     The proof process is omitted here. Please refer to the full version of this paper in \hl{[]}.
% \end{proof}
\begin{pf}
    Assume that the set of neighboring sensors of sensor $i$ has been approximately divided into $m_i$ subsets,  the expected number of attacked sensors and the balance of attack intensity in each subset will be analyzed.

First, let's analyze the expected number of attacked sensors in each subset. Since the set of attacked sensors changes randomly at each moment, we estimate the mathematical expectation of the total number of attacked sensors over the entire period using the Monte Carlo method. The probability of a sensor being selected as an attack target is $q_i / |\mathcal{N}_i|$. Therefore, the expected number of attacked sensors in each subset is given by $\mathbb{E}[q_{i,\,g_i}] = |\mathcal{N}_{i,\,g_i}| \times \left( \frac{q_i}{|\mathcal{N}_i|} \right) \approx \frac{|\mathcal{N}_i|}{m_i} \times \frac{q_i}{|\mathcal{N}_i|} = \left\lfloor \frac{q_i}{m_i} \right\rfloor$.
%\[
%\mathbb{E}[q_{i,\,g_i}] = |\mathcal{N}_{i,\,g_i}| \times \left( \frac{q_i}{|\mathcal{N}_i|} \right) \approx \frac{|\mathcal{N}_i|}{m_i} \times \frac{q_i}{|\mathcal{N}_i|} = \left\lfloor \frac{q_i}{m_i} \right\rfloor.
%\]

Next, we consider the balance of attack intensity. In the literature \cite{Suo2024attack}, the average malicious disturbance power of the attack signal $z_{ij}(k)$ is defined as the average of the sum of the squared values of the attack signal $z_{ij}(k)$ on the communication link between sensor $j$ and sensor $i$ over the entire time period. However, since the attacked communication links vary dynamically at each moment, and the same link may experience different attacks at different moment, the definition in \cite{Suo2024attack} is not applicable in this paper.

Therefore, this paper redefines the average malicious disturbance power from the attacker's perspective, rather than from the specific communication link's perspective. For the $l$-th attacker surrounding sensor $i$, where $l=1:q_i$, the average malicious disturbance power $\phi_{i,\,l}$ of the injected attack signal $z_{i,\,l}$ is defined as $\phi_{i,\,l} = \lim_{T \to \infty} \frac{1}{T} \sum_{k=1}^T \left( z_{i,\,l}(k) \right)^2$.
%\begin{equation}
%\phi_{i,\,l} = \lim_{T \to \infty} \frac{1}{T} \sum_{k=1}^T \left( z_{i,\,l}(k) \right)^2.
%\end{equation}
Based on the previous analysis, if the attacked sensors are completely random at each moment, the expected number of attacked sensors in each subset is $\left\lfloor q_i/{m_i} \right\rfloor$. Based on Assumption \ref{chapter2:attack_assum}, the ratio of the average malicious disturbance power of each subset to the total malicious disturbance power of all attacks is $\left\lfloor q_i/m_i \right\rfloor/q_i= 1/m_i
$.
This indicates that the attack intensity in each subset is balanced. This completes the proof.
\end{pf}

Then, a basic numerical simulation is conducted to verify that the attacker satisfies Lemma~\ref{lem41}. According to the set partitioning strategy in Algorithm~\ref{alg:3-4}, the 100 sensors are divided into four subsets, where sensors 1–25 form subset 1, sensors 26–50 form subset 2, sensors 51–75 form subset 3, and sensors 76–100 form subset 4. Without considering specific attack types, the attacker randomly selects 40 sensors at each time step. The statistical results over the entire time horizon are summarized in Table~\ref{tab:attack_stats}, showing that the average number of sensors attacked in each subset is close to the ideal value of 10.

When specific attack types are considered, the attack intensity distribution across subsets can also be evaluated. As shown in Table~\ref{tab:attack_stats}, the ratio of the attack intensity in each subset to the total attack signal intensity ideally equals 0.25. The numerical results demonstrate that the observed intensity ratios fluctuate slightly around this theoretical value, for both unstealthy and stealthy attacks.

%Then, a basic numerical simulation is utilized to verify that the attacker satisfies Lemma \ref{lem41}. Based on the set partitioning strategy in Algorithm \ref{alg:3-4}, the 100 sensors are divided into four subsets: sensors 1–25 form subset 1, sensors 26–50 form subset 2, sensors 51–75 form subset 3, and sensors 76–100 form subset 4. Without considering specific attack types, at each time step, the attacker randomly selects 40 sensors for attack. According to statistics, During the entire time period, the average number of sensors attacked in each subset was 9.83, 9.97, 10.04, and 10.16, respectively, which is close to the ideal average of 10.
%
%When specific attack types are considered, the attack intensity distribution for each subset can be further examined. The ratio of the attack intensity in each subset to the total attack signal intensity is ideally 0.25. The results show that the intensity ratios fluctuate around this ideal value. In one case, the mean attack intensity ratios for the four subsets are 0.2497, 0.2500, 0.2504, and 0.2499, while in another case the corresponding mean ratios are 0.2507, 0.2490, 0.2499, and 0.2504.

\begin{table}[htb]
	\centering
	\caption{Attack statistics across subsets compared with the ideal values}
	\begin{tabular}{c|cccc|c}
		\hline
		& Subset 1 & Subset 2 & Subset 3 & Subset 4 & Ideal \\
		\hline
		Average number of attacked sensors & 9.83 & 9.97 & 10.04 & 10.16 & 10 \\
		\hline
		Mean attack intensity ratio of unstealthy attacks & 0.2497 & 0.2500 & 0.2504 & 0.2499 & 0.25 \\
		Mean attack intensity ratio of stealthy attacks & 0.2507 & 0.2490 & 0.2499 & 0.2504 & 0.25 \\
		\hline
	\end{tabular}
	\label{tab:attack_stats}
\end{table}

\section{APPENDIX B: Distributed Attack Detection Scheduling Algorithm}

The ADS algorithm in literature \cite{Suo2024Security} can be extended to the D-ADS algorithm, as shown in Algorithm \ref{alg:3-1}. The main difference between the D-ADS and ADS algorithms lies in the sensor selection method, specifically whether it uses centralized selection or distributed two-stage sensor selection. For the $l$-th selection at time step $k$, a malicious information is first chosen from each subset (Stage 1 of sensor selection) and added to the set $\mathcal{S}^{(l)}_k$. Then, a sensor is randomly selected from $\mathcal{S}^{(l)}_k$ as the result of the $l$-th selection (Stage 2 of sensor selection). The algorithm terminates when $|\mathcal{A}_{i,\,k}^{(l)}|=q_i$. 

It should be noted that in steps 4-7, the gains of all remaining sensors $j \in \mathcal{N}_i \backslash \mathcal{A}_{i,\,k}^{(l-1)}$ are updated. However, with the improved partitioning strategy, only the gains of one subset of sensors need to be updated here. The meanings of the parameters in Algorithm \ref{alg:3-1} are summarized in TABLE \ref{tab:distributed parameter}\footnote{Other variable symbols that are not explained here are  the same  as literature \cite{Suo2024Security}}.

\begin{algorithm}[t]  
%\setstretch{1.3}
\renewcommand{\thealgorithm}{3.3}
	\caption{Distributed Attack Detection Scheduling Algorithm}
	\label{alg:3-1}  
 
	\begin{algorithmic}[1]  
		\Require  
		Neighbor set $\mathcal{N}_i$ of sensor $i$, maximum number of attacked neighboring sensors $q_i$, historical information values $W_{kj}$, $j\in\mathcal {N}_i$.  
		\Ensure  
		Attacked sensor set $\mathcal{A}_{i,\,k}$ at time $k$ and $\mathcal{A}_{i,\,k,\,g_i}=\mathcal{A}_{i,\,k,\,g_i}^{(l)}$, where $g_i=1:m_i$, $k=1,2,...,T$.
		
		\State Initialize weight vector $\omega_{k}^{(l)}=[\omega_{kj}^{(l)}]_{j\in\mathcal{N}_{i}}$, 
  where each element $\omega _{kj}=1$, and set $\mathcal{A}_{i,\,k,\,g_i}^{(0)}=\emptyset$ for $p=1:m_i$, $k=1:T$, $l=l_{g_i}=1$.

		\While{$l<q_i$}
		\State Set $\mathcal{S}_{k}^{(l)}=\mathcal{A}_{i,\,k}^{(l)}=\emptyset$.
        \For {all $j\in \mathcal{N}_i\backslash\mathcal{A}_{i,\,k}^{(l-1)}$} 
        \Statex \quad\quad\quad $\%$ With optimized grouping strategies, only a subset of sensors' gains needs to be updated here.
		\State Compute $G_{kj}^{(l)} = f_k( \mathcal{A}_{i,\,k}^{(l-1)} )-f_k( \mathcal{A}_{i,\,k}^{(l-1)}\cup \left\{ j \right\} ) $.
    \State Update $w_{k}^{(l)}$ as $w_{kj}^{(l)}= w_{k,\,j}^{(l-1)}e^{ -G_{kj}^{(l)}}$.
  \EndFor
		
		\For{$g_i=1:m_i$}
		\State Set $w_{g_i,\,k}^{(l_{g_i})}=\left[ w_{kj} \right] _{j\in \mathcal{N} _{i,\,g_i}\backslash \mathcal{A} _{i,\,k,\,g_i}^{(l_{g_i}-1)}}$.
		\State Compute $p_{g_i,\,k}^{(l_{g_i})}=w_{g_i,\,k}^{(l_{g_i})}/\| w_{g_i,\,k}^{(l_{g_i})}\|_1$.
		\State Select an element $j^{(l_{g_i})}_{g_i,select}$ based on the distribution vector $p_{g_i,k}^{(l_{g_i})}$. $\%$ Stage 1 sensor selection.
		\State Obtain $\mathcal{S}_{k}^{(l)}=\mathcal{S}_{k}^{(l)}\cup \{j^{(l_{g_i})}_{g_i,\,select}\}$.

		\EndFor

		\State Randomly select an element $j_{g_i^s,\,select}^{(l_{g_i^s})}$ from $\mathcal{S}_{k}^{(l)}$. $\%$ Stage 2 sensor selection.
		\State Obtain $\mathcal{A}_{i,\,k}^{(l)}=\mathcal{A}_{i,\,k}^{(l-1)}\cup \{j_{g_i^s,\,select}^{(l_{g_i^s})}\}$.
		\State Obtain $\mathcal{A}_{i,\,k,\,g_i^s}^{(l_{g_i^s})}=\mathcal{A}_{i,\,k,\,g_i^s}^{(l_{g_i^s}-1)}\cup \{j^{(l_{g_i^s})}_{{g_i^s},\,select}\}$.
		\State Update $l=l+1$.
		\State Update $l_{g_i^s}=l_{g_i^s}+1$.
		\EndWhile

		\State \Return $\mathcal{A}_{i,k}$ and $\mathcal{A}_{i,\,k,\,g_i}=\mathcal{A}_{i,\,k,\,g_i}^{(l)}$, where $g_i=1:m_i$.
	\end{algorithmic}  
\end{algorithm}

\begin{table*}[ht]
%\small
\renewcommand{\arraystretch}{1.3}
\centering
\caption{Summary of variables in Algorithm \ref{alg:3-1}}
\begin{tabular*}{0.9\textwidth}{@{\extracolsep{\fill}}cc}
\toprule
Parameter & Description \\
\midrule
$\omega_{g_i,\,k}^{(l_{g_i})}$ & The weight vector to be updated at the $l_{g_i}$-th selection for the $g_i$-th subset at moment $k$ \\

$\mathcal{A}_{i,\,k,\,g_i^s}^{l_{g_i^s}}$ & The set of attacked sensors after the $l_{g_i^s}$-th selection for the $g_i^s$-th subset at moment $k$ \\

$p_{g_i,\,k}^{(l_{g_i})}$ & The distribution proportion vector at the $l_{g_i}$-th selection for the $g_i^s$-th subset at moment $k$ \\

$\mathcal{S}^{(l)}_{k}$ & The set of $g_i$ sensors selected from all subsets at the $l$-th selection at moment $k$ \\

$j_{g_i^s,\,select}^{l_{g_i^s}}$ & The sensor selected in the $l_{g_i^s}$-th selection for the $g_i^s$-th subset \\

$g_i^s$ & The subset to which the sensor $j_{g_i^s,\,select}^{l_{g_i^s}}$ selected from $\mathcal{S}^{(l)}_{k}$ belongs \\

\bottomrule
\end{tabular*}
\label{tab:distributed parameter}
\end{table*}

Next, the performance relationship between the D-ADS algorithm in this paper and the ADS algorithm in literature \cite{Suo2024Security} is analyzed. Both algorithms aim to select the set of malicious information that have the greatest impact on the objective function (\ref{f_A}). Therefore, ideally, their performance is consistent, meaning that $\mathcal{A}_{i,\,k} = \cup^{m_i}_{g_i=1}{\mathcal{A}_{i,\,k,\,g_i}}
$.

\begin{lem}
For $g_i\in\{1,\,...,\,m_i\}$, $l\in\{0,\,1,\,...,\,q_i\}$, define $\delta_{l,\,g_i}$ as $\delta _{l,\,g_i}=\sum_{k=1}^T{(}\frac{1}{m_i}f_k(\mathcal{A} _{i,\,k}^{*})-f_k(\mathcal{A} _{i,\,k,\,g_i}^{(l_{g_i})}))$
, where $\mathcal{A}_{i,\,k,\,g_i}^ {(l_{g_i})}$ represents the set of attacked sensors after the $l_{g_i}$-th selection of the $g_i$-th subset at moment $k$.
Then, the relationship in literature \cite{Suo2024Security} can be transformed as $\sum^{m_i}_{g_i=1}\delta_{l+1,\,g_i} - (1-\frac{1}{q_i})^{l+1}\sum^{m_i}_{g_i=1}\delta_{0,\,g_i}\le \frac{m_i}{q_i}\sum_{j=1}^{l+1}(1-\frac{1}{q_i})^{l+1-j}  B_i^{(j)}$.
% \begin{equation}\label{relationship}		\sum^{m_i}_{g_i=1}\delta_{l+1,g_i} - (1-\frac{1}{q_i})^{l+1}\sum^{m_i}_{g_i=1}\delta_{0,g_i}\le \frac{m_i}{q_i}\sum_{j=1}^{l+1}(1-\frac{1}{q_i})^{l+1-j}  B_i^{(j)}.
% \end{equation}
The parameter $B_i^{(l)}$ is described in detail in the proof. At this time, the Lemma 3.2 in literature\cite{Suo2024Security} is extended to the distributed case.
\end{lem}
% \begin{proof}
%     The proof process is omitted here. Please refer to the full version of this paper in \hl{[]}.
% \end{proof}
\begin{pf}
For the D-ADS algorithm, by summing $\delta_{l,\,g_i}$ over all $m_i$ subsets and taking the average, we obtain
    \begin{eqnarray}\label{delta_lp_relax}
% &=& \frac{1}{m_i}\sum^T_{k=1}(f_k(\mathcal{A}_{i,k}^{*})-\sum^{m_i}_{g_i=1}f_k(\mathcal{A}_{i,k,g_i}^{(l_{g_i})}))) \nonumber \\
\frac{1}{m_i}\sum^{m_i}_{g_i=1}\delta_{l,g_i}&\le&\frac{1}{m_i}\sum^T_{k=1}(\sum^{m_i}_{g_i=1}(f_k(\mathcal{A}_{i,\,k,\,g_i}^{*})-f_k(\mathcal{A}_{i,\,k,\,g_i}^{(l_{g_i})}))) \nonumber \\
&\le& \frac{1}{m_i}\sum_{k=1}^T (\sum^{m_i}_{g_i=1} \sum_{j\in \mathcal{N}_{i,\,k,\,g_i}\backslash\mathcal{A}_{i,\,k,\,g_i}^*} ( f_k(\mathcal{A}_{i,\,k,\,g_i}^{(l_{g_i})}\cup \{j_{kl}^*\})-f_k(\mathcal{A}_{i,\,k,\,g_i}^{(l_{g_i})}) )) \nonumber \\
&=& \frac{1}{m_i}\sum_{k=1}^T (  \sum^{m_i}_{g_i=1} ( -\sum_{j \in \mathcal{N}_{i,\,k,\,g_i}\backslash\mathcal{A}_{i,\,k,\,g_i}^*}G_{kj}^{(l_{g_i}+1)} ) ),
\end{eqnarray}
!where the first inequality follows from $f_k(\mathcal{A}_{i,\,k}^{*}) \le \sum^{m_i}_{g_i=1}f_k(\mathcal{A}_{i,\,k,\,g_i}^{(l_{g_i})})$, the second inequality follows from the submodularity of $f_k$, and the third equality follows from the definition of $G_{kj}^{(l+1)}$.  

Based on Lemma 3 in literature \cite{matsuoka2021tracking}, as well as the fact that $\delta_l \le \sum^{m_i}_{g_i=1} \delta_{l,\,g_i}$ and $\delta_{l,g_i} \ge \delta_{l+1,\,g_i}$, we have
\begin{equation}
\frac{1-q_i}{m_i}\sum^{m_i}_{g_i=1}\delta_{l,\,g_i} \le -\frac{q_i}{m_i} \left( \sum^{m_i}_{g_i=1} \delta_{l+1,\,g_i} \right) + B^{(l+1)}_i,
\end{equation}
where
\begin{equation}
B^{(l)}_i = \sum_{j=1}^{q_i}\sum_{k=1}^T \left( \frac{1}{m_i}\sum^{m_i}_{g_i=1} G_{k,\,g_i}^{(l_{g_i})} p_{g_i,\,k}^{(l_{g_i})} - G_{kj_{kl}^{*}}^{(l)} \right),
\end{equation}
and $G_{k,\,g_i}^{(l_{g_i})} = [ G_{kj}^{(l)} ]_{j \in \mathcal{N}_{i,\,k,\,g_i} \backslash \mathcal{A}_{i,\,k,\,g_i}^{(l_{g_i}-1)}}$, with $G_{kj_{kl}^{*}}^{(l)}$ denoting the gain of the optimal sensor $j_{kl}^{*}$ at time $k$ in the $l$-th selection. 

Thus, for all $l \in \{0, 1, ..., q_i\}$, the inequality 
\begin{equation}
\sum^{m_i}_{g_i=1} \delta_{l+1,\,g_i} - \left(1 - \frac{1}{q_i} \right) \sum^{m_i}_{g_i=1} \delta_{l,\,g_i} \le \frac{m_i}{q_i} B_i^{(l+1)}
\end{equation}
holds. After iterating, we obtain
\begin{equation}\label{relationship}
   \sum^{m_i}_{g_i=1} \delta_{l+1,\,g_i} - \left( 1 - \frac{1}{q_i} \right)^{l+1} \sum^{m_i}_{g_i=1} \delta_{0,g_i} \le \frac{m_i}{q_i} \sum_{j=1}^{l+1} \left( 1 - \frac{1}{q_i} \right)^{l+1-j} B_i^{(j)}. 
\end{equation}

Thus, Lemma 1 in literature \cite{Suo2024Security} has been extended to the distributed case.
This completes the proof.
\end{pf}

\begin{thm}\label{thm42}
    For the dynamic attack strategy in the Definition \ref{semi-dynamic attack defn}, the proposed D-ADS algorithm can ensure that the theoretical lower bound of the average optimization rate of malicious information selection for any subset is $1-1/e$, and the error expectation is bounded.
\end{thm}
% \begin{proof}
%     The proof process is omitted here. Please refer to the full version of this paper in \hl{[]}.
% \end{proof}
\begin{pf}
To prove that the theoretical lower bound of the average optimization rate over the entire time period is 
$1 - 1/e$, 
we need to show that the expectation of the error 
\begin{equation}\label{chapter3:difference_between_real_and_predict}
\mathbb{E} \left[ \left( 1 - \frac{1}{e} \right) \sum_{k=1}^T{\frac{1}{m_i} f_k\left( \mathcal{A}_{i,\,k}^{*} \right)} - \sum_{k=1}^T{f_k( \mathcal{A}_{i,\,k,\,g_i} )} \right]  
\end{equation}
is bounded, where $f_k\left( \mathcal{A}_{i,\,k,\,g_i} \right)$ represents the submodular function value for the selected malicious information set in subset $g_i$ at time $k$.

Essentially, the expectation of the error represents the expected difference between the objective function value of the selected malicious information set and $1 - 1/e$ times the optimal malicious information set's objective function value. Therefore, the larger the objective function value of a subset, the greater the upper bound of the error expectation. Consequently, the subset with the largest objective function value has the highest upper bound for error expectation, as shown in equation (\ref{chapter3:Theorem32_25})
\begin{eqnarray}\label{chapter3:Theorem32_25}
&&( 1 - \frac{1}{e} ) \sum_{k=1}^T{\frac{1}{m_i} f_k\left( \mathcal{A}_{i,\,k}^{*} \right)} - \sum_{k=1}^T{\max_{g_i} f_k( \mathcal{A}_{i,\,k,\,g_i} )} \nonumber\\ 
&\le& \frac{1}{m_i} \left[ (1 - \frac{1}{e}) \sum_{k=1}^T{ f_k\left( \mathcal{A}_{i,\,k}^{*} \right)} - \sum_{k=1}^T{ f_k( \mathcal{A}_{i,\,k} )} \right] \\
&\le&\frac{1}{m_i} \left[ (1 - (1 - \frac{1}{q_i})^{q_i})\sum_{k=1}^T{f_k\left( \mathcal{A}_{i,\,k}^{*} \right)} - \sum_{k=1}^T{f_k\left( \mathcal{A}_{i,\,k} \right)} \right],\nonumber
\end{eqnarray}
where the first inequality follows from literature \cite{mirzasoleiman2016distributed}, which states that $\max_{g_i} f_k(\mathcal{A}_{i,\,k,\,g_i}) \ge \frac{1}{m_i} f_k(\mathcal{A}_{i,\,k})$, and the second inequality follows from $(1 - 1/k)^{k} \le 1/e$. 

It is important to note that $\sum_{k=1}^T 1/{m_i} \cdot f_k\left( \mathcal{A}_{i,\,k}^{*}\right)$ does not imply that the objective function of each subset satisfies $\sum_{k=1}^T 1/{m_i} \cdot f_k\left( \mathcal{A}_{i,\,k}^{*}\right) = \sum_{k=1}^T f_k\left( \mathcal{A}_{i,\,k,\,g_i}^{*}\right)$. Instead, it follows from the assumption that the impact of attack signals on each subset is balanced over the entire time period.

Based on the Lemma 3 from literature \cite{matsuoka2021tracking} and the definition $\delta_l = \sum_{k=1}^T \left( f_k( \mathcal{A}_{i,\,k}^{*} ) - f_k ( \mathcal{A}_{i,\,k}^{(l)} ) \right)$ for $l=1:q_i$, we obtain that,
$\sum_{k=1}^T{f_k\left( \mathcal{A}_{i,\,k}^{*} \right)} - \sum_{k=1}^T{f_k \left( \mathcal{A}_{i,\,k} \right)} + (1 - (1 - \frac{1}{q_i})^{q_i})\sum_{k=1}^T{(f_k( \mathcal {A}_{i,\,k}^{(0)} ) - f_k( \mathcal{A}_{i,\,k}^{*} ))} \le \delta_{q_i} - (1 - \frac {1}{q_i})^{q_i} \delta_0 \le \sum^{m_i}_{g_i=1} \delta_{l,\,g_i} - (1 - \frac{1}{q_i})^{q_i} \sum^{m_i}_{g_i=1} \delta_{0,\,g_i}
$, where the inequality $\delta_{q_i} \le \sum^{m_i}_{g_i=1} \delta_{l,\,g_i}$ follows from the diminishing marginal returns property of the submodular function, and the initial condition satisfies $\delta_0 = \sum_{g_i=1}^{m_i} \delta_{0,\,g_i}$. Combining equation (\ref{relationship}) with the above relation, equation (\ref{chapter3:Theorem32_25}) can be rewritten as
% \begin{equation}
% ( 1 - \frac{1}{e} ) \sum_{k=1}^T{\frac{1}{m_i} f_k\left( \mathcal{A}_{i,k}^{*} \right)} - \sum_{k=1}^T{\max_{g_i} f_k( \mathcal{A}_{i,k,g_i} )}
% \le \frac{m_i}{q_i} \sum_{j=1}^{q_i} (1 - \frac{1}{q_i})^{q_i-j} \frac{B_i^{(j)}}{m_i}.
% \end{equation}
\begin{eqnarray}
   &&( 1 - \frac{1}{e} ) \sum_{k=1}^T{\frac{1}{m_i} f_k\left( \mathcal{A}_{i,\,k}^{*} \right)} - \sum_{k=1}^T{\max_{g_i} f_k( \mathcal{A}_{i,\,k,\,g_i} )}
\le \frac{m_i}{q_i} \sum_{j=1}^{q_i} (1 - \frac{1}{q_i})^{q_i-j} \frac{B_i^{(j)}}{m_i}. 
\end{eqnarray}

Thus, to prove that equation (\ref{chapter3:Theorem32_25}) is bounded, it suffices to show that $\mathbb{E} [B_i^{(l)}/m_i]$ is bounded
\begin{eqnarray}\label{equ4.8}
\mathbb{E} \left[ \frac{B_i^{(l)}}{m_i} \right]&=& \frac{1}{m_i} \sum_{j=1}^{q_i} \mathbb{E} \left[ \sum_{k=1}^T \left( \frac{1}{m_i} \sum^{m_i}_{g_i=1} G_{k,\,g_i}^{( l_{g_i} )} p_{g_i,\,k}^{( l_{g_i} )} - G_{kj_{kl}^{*}}^{\left( l \right)} \right) \right] 
\\
&\le& \frac{1}{m_i} \sum_{j=1}^{q_i} \mathbb{E} \left[ \sum_{k=1}^T ( G_{k}^{\left( l \right)} p_{k}^{\left( l \right)} - G_{kj_{kl}^{*}}^{\left( l \right)} ) \right]  
\nonumber\\
&\le& \frac{2\,q_i}{m_i} \sqrt{{( 2\varDelta _T\log ( |\mathcal{N}_i|T ) + T ( 2\log ( |\mathcal{N}_i|T ) + \log ( T ) ) )}},\nonumber
\end{eqnarray}
where the first inequality follows from the arithmetic mean-geometric mean inequality, and the second follows from Theorem 2 in literature \cite{Suo2024Security}.

Thus, the expectation of the error in equation (\ref{chapter3:difference_between_real_and_predict}) is bounded
\begin{eqnarray}\label{chapter3:dynamic_expextation}
   && \mathbb{E} \left[ \left( 1 - \frac{1}{e} \right) \sum_{k=1}^T{\frac{1}{m_i} f_k\left( \mathcal{A}_{i,\,k}^{*} \right)} - \sum_{k=1}^T{f_k( \mathcal{A}_{i,\,k,\,g_i} )} \right]  
\le \tilde{\mathcal{O}} \left( \frac{q_i}{m_i} \sqrt{3T+2\varDelta _T} \right),
\end{eqnarray}
where $\mathbb{E} [\cdot]$ represents expectation, and $\tilde{\mathcal{O}}$ hides logarithmic terms.
This completes the proof.
\end{pf}

\printcredits

\section*{Acknowledgments}
The authors note that this paper has also been made available as a preprint on arXiv \citep{suo2025efficient}.

\section*{Declaration of Interest}
The authors declare that they have no known competing financial interests or personal relationships that could have appeared to influence the work reported in this paper. Generative AI and AI-assisted technologies are only applied to the writing process to improve the readability and language of the manuscript.

\bibliographystyle{cas-model2-names}

\bibliography{ref}


%Author biography without author photo.
%Author biography. Author biography. Author biography.
%Author biography. Author biography. Author biography.
%Author biography. Author biography. Author biography.
%Author biography. Author biography. Author biography.
%Author biography. Author biography. Author biography.
%Author biography. Author biography. Author biography.
%Author biography. Author biography. Author biography.
%Author biography. Author biography. Author biography.
%Author biography. Author biography. Author biography.

%\bio{figs/pic1}
%Author biography with author photo.
%Author biography. Author biography. Author biography.
%Author biography. Author biography. Author biography.
%Author biography. Author biography. Author biography.
%Author biography. Author biography. Author biography.
%Author biography. Author biography. Author biography.
%Author biography. Author biography. Author biography.
%Author biography. Author biography. Author biography.
%Author biography. Author biography. Author biography.
%Author biography. Author biography. Author biography.
%\bio{figs/pic1}
%Author biography with author photo.
%Author biography. Author biography. Author biography.
%Author biography. Author biography. Author biography.
%Author biography. Author biography. Author biography.
%Author biography. Author biography. Author biography.

%\end{sloppypar}
\end{document}